# Spectrophotometric Modeling and Mapping of Ceres

Jian-Yang Li (李荐扬) [a,*], Stefan E. Schröder [b], Stefano Mottola [b], Andreas Nathues [c], Julie C. Castillo-Rogez [d], Norbert Schorghofer [a], David A. Williams [e], Mauro Ciarniello [f], Andrea Longobardo [f], Carol A. Raymond [d], Christopher T. Russell [g]

[a] Planetary Science Institute, Tucson, AZ 85719, USA
[b] Deutsches Zentrum für Luft- und Raumfahrt (DLR), 12489 Berlin, Germany
[c] Max Planck Institute for Solar System Research, Justus-von-Liebig-Weg 3, D-37077, Goettingen, Germany
[d] Jet Propulsion Laboratory (JPL), California Institute of Technology, Pasadena, CA 91109-8099, USA
[e] School of Earth and Space Exploration, Arizona State University, Tempe, AZ 85287, USA
[f] Istituto di Astrofisica e Planetologia Spaziali, Istituto Nazionale di Astrofisica (INAF), 00133 Rome, Italy
[g] Institute of Geophysics and Planetary Physics (IGPP), University of California, Los Angeles, CA 90095-1567, USA

* Corresponding author:

Jian-Yang Li
Planetary Science Institute
1700 E. Ft. Lowell Rd., Suite 106
Tucson, AZ 85719, USA
jyli@psi.edu
+1 571-488-9999

**Abstract:**

We report a comprehensive analysis of the global spectrophotometric properties of Ceres using the images collected by the Dawn Framing Camera through seven color filters from April to June 2015 during the RC3 (rotational characterization 3) and Survey mission phases. We derived the Hapke model parameters for all color filters. The single-scattering albedo of Ceres at 555 nm wavelength is 0.14±0.04, the geometric albedo is 0.096±0.006, and the Bond albedo is 0.037±0.002. The asymmetry factors calculated from the best-fit two-term Henyey-Greenstein (HG) single-particle phase function (SPPF) show a weak wavelength dependence from -0.04 at 438 nm increasing to 0.002 at >900 nm, suggesting that the phase reddening of Ceres is dominated by single-particle scattering rather than multiple scattering or small-scale surface roughness. The Hapke roughness parameter of Ceres is derived to be 20º±6º, with no wavelength dependence. The phase function of Ceres presents appreciably strong scattering around 90º phase angle that cannot be fitted with a single-term HG SPPF, suggesting possible stronger forward scattering component than other asteroids previously analyzed with spacecraft data. We speculate that such a scattering characteristic of Ceres might be related to its ubiquitous distribution of phyllosilicates and high abundance of carbonates on the surface. We further grouped the reflectance data into a 1º latitude-longitude grid over the surface of Ceres, and fitted each grid independently with both empirical models and the Hapke model to study the spatial variations of photometric properties. Our derived albedo maps and color maps are consistent with previous studies [Nathues, A., et al., 2016, Planet. Space Sci. 134, 122-127; Schröder, S.E., et al., 2017, Icarus 288, 201-225]. The SPPF over the surface of Ceres shows an overall correlation with albedo distribution, where lower albedo is mostly associated with stronger backscattering and vice versa, consistent with the general trend among asteroids. On the other hand, the Hapke roughness parameter does not vary much across the surface of Ceres, except for the ancient Vendimia Planitia region that is associated with a slightly higher roughness. Furthermore, the spatial distributions of the SPPF and the Hapke roughness do not depend on wavelength. Based on the wavelength dependence of the SPPF of Ceres, we hypothesize that the regolith grains on Ceres either contain a considerable fraction of μm-sized or smaller particles, or are strongly affected by internal scatterers of this size.

*Keywords:* Asteroid Ceres; Asteroids, surfaces; Photometry; Spectrophotometry

## 1. Introduction

In orbit around Ceres since March 2015, NASA's Dawn spacecraft has collected a large amount of multispectral imaging data with the onboard Framing Camera (FC) in the visible wavelength, allowing for a detailed study of the photometric properties of Ceres. This article focuses on the analysis of the global spectrophotometric properties of Ceres, as well as a mapping of photometric properties through modeling parameters using the FC data.

Ceres has been shown to be an active world that is strongly affected by water (ice and/or hydrates) on its surface and crust (Sizemore et al., submitted). The prevalent distribution of ammoniated phyllosilicates suggests a widespread aqueous alteration in Ceres' interior (De Sanctis et al. 2015; Ammannito et al. 2016). Abundant hydrogen most likely reveals a global distribution of water ice and/or hydration beneath the surface, more abundant at mid- to high-latitude (Prettyman et al., 2017). A few kilometer-sized water ice patches are identified in isolated regions associated with young craters (Combe et al. 2016). Pitted terrains (Sizemore et al., 2017) and flow-like geomorphological features (Schmidt et al., 2017) are additional indicators of abundant water ice in the shallow subsurface. Although conflicting evidence exists about the amount of water ice contained in Ceres' crust (Hiesinger et al., 2016; Bland et al., 2016), it is clear that the present physical properties on the surface of Ceres are strongly affected by water ice, and are very different from "dry" asteroids such as Vesta (cf. Keil, 2002).

Before Dawn's observations of Ceres, the photometric properties of Ceres had been studied exclusively from ground-based observations of its phase function (see a review in Reddy et al., 2015). The historical phase function data of Ceres appear to be consistent with an IAU H-G model with $H$=3.34 and $G$=0.10 to 0.12 (Tedesco, 1989; Tedesco et al., 2002), and with a Hapke model[1] having a single-scattering albedo (SSA), $w$=0.070, an asymmetry factor of the single-term Henyey-Greenstein (1pHG) single-particle phase function (SPPF), $\xi$=-0.40, an amplitude $B_0$=1.6 and a width $h$=0.06 of the shadow-hiding opposition effect, and an assumed macroscopic roughness $\bar{\theta}$ of 20º (Helfenstein and Veverka, 1989). Reddy et al. (2015) reported ground-based observations of Ceres with a spare set of FC color filters (Sierks et al., 2011), and a set of Hapke parameters of $w$=0.083, $\xi$=-0.37, $B_0$=2.0, $h$=0.036, with an assumed roughness of 20º. Li et al. (2006) used images from the Hubble Space Telescope (HST) to perform a photometric modeling with the Hapke model, although they had to adopt $\xi$=-0.40 based on Helfenstein and Veverka (1989) because of the small range of about 2º in the phase angles of their data. They derived an SSA of 0.070 and a geometric albedo of 0.092 at 555 nm wavelength. The high roughness of 44º that Li et al. (2006) reported is likely a modeling artifact (see Section 4.3).

Schröder et al. (2017) present a comprehensive study of the photometric properties of Ceres based on FC images. They reported that the "disk-function" of Ceres, which describes the dependence of surface reflectance on local topography (incidence angle, $i$, and emission angle, $e$), can be described by both the Akimov model (Shkuratov et al., 2011) and the Hapke model equally well. They found a set of Hapke parameters based on a two-parameter Henyey-Greenstein (2pHG) function, with parameters $w$=0.11, $B_0$=4.0, $h$=0.02, $\bar{\theta}$=22º, $b$=0.30, and $c$=0.65[2], but their values of $h$, $b$, and $c$ were all manually chosen. After correcting for disk-function, Schröder et al. (2017) used RC3 data to map out the normal albedo $A_N$ and phase slope $\nu$ of Ceres by fitting the equigonal albedo $A_{eq}(\alpha)$ at each latitude-longitude position on the whole surface with a simple exponential model, $A_{eq}(\alpha) = A_N \exp(-\nu\alpha)$, where $\alpha$ is phase angle. While the albedo map derived this way is consistent with that derived by the traditional photometric correction showing many bright features associated with geologically young craters, the phase slope map appears to be mostly featureless on a global scale, with some slight correlation with the geological settings of craters on local scales. This contrasts with Vesta, where a clear correlation between the phase slope and geological settings is evident and has been interpreted as roughness driven by geological age (Schröder et al., 2013a).

---

[1] The symbols of all Hapke parameters from the literature have been adopted following the formula, parameters and symbols as described in Section 3.1

[2] Schröder et al. (2017) adopted a 2pHG that has its $c$ parameter equivalent to -$c$ in the form we adopted. The value of 0.65 we quoted here has been converted to our form of the 2pHG. Similarly, for all values of $b$ and $c$ for the 2pHG

Ciarniello et al. (2017) reported their comprehensive photometric analysis of Ceres with the Hapke model in both the visible and near-infrared wavelengths using the Dawn visible and infrared spectrometer (VIR, De Sanctis et al. 2011) data. At 550 nm wavelength, assuming $B_0$=1.6 and $h$=0.06, they fitted a set of photometric parameters $w$=0.14±0.02, $\bar{\theta}$=29°±6°, and derived an asymmetry factor $\xi = -bc = -0.11\pm0.08$ from their best-fit 2pHG parameters. This model is mostly consistent with the model derived by Schröder et al. (2017), although some differences exist, which could arise from their different treatments of the opposition, as well as the slightly different approaches in model fitting. Phase reddening is observed throughout visible to near-infrared wavelengths.

Longobardo et al. (2018) studied the photometric properties of Ceres with an empirical approach. They concluded that Akimov disk-function model is the best among the models they tested to fit the Dawn VIR data, and confirmed an overall uniformity across the Ceres surface and obtained albedo maps in good agreement with previous work (Ciarniello et al., 2017; Schröder et al., 2017), with the exception of the bright faculae inside Occator crater, characterized by a phase slope steeper than expected for their high albedo. This has been ascribed to a higher roughness of this region. Moreover, they found that phase reddening is weaker or absent in correspondence of carbonate enrichments.

In April 2017, Dawn collected data at phase angles 0° - 7° for the purpose of studying the opposition effect of Ceres' regolith, particularly in the extremely bright Cerealia Facula. Schröder et al. (in press) analyzed the data with primarily an empirical approach, and reported that the opposition effect of Ceres is typical for a C-type asteroid. The characteristics of the opposition effect of Ceres do not systematically vary with wavelength, and they do not vary across the studied region between -60° and +30° in latitude and 160° to 280° in longitude, with an exception in the fresh ejecta of Azacca crater that displays an enhancement at phase angles <0.5°. The broadband visible geometric albedo of Ceres is precisely measured to be 0.094±0.005 at opposition. However, the Hapke model failed to converge to a reasonable set of parameters for the opposition effect.

The goals of our study are: 1. To derive a set of global Hapke photometric model parameters in all color filters to characterize the light scattering behaviors of Ceres' surface; 2. To provide maps of photometric models in all color wavelengths in order to understand the variations of photometric properties across the whole surface of Ceres. We will present the data that we used, as well as the processing and reduction in Section 2, describe the details of the models in Section

---

functions we quoted in this manuscript from the literature, we have converted them to be consistent with our form of 2pHG.

3.  The results of global photometric modeling will be reported in Section 4, and the photometric model mapping results in Section 5.  Section 6 discusses the implications of our results.  Section 7 summarizes the major findings and conclusions of our study.

## 2.  Dataset

*2.1. Data and Calibration*

We used images collected by the FC (Sierks et al., 2011) in this study.  The FC has two identically manufactured cameras, and FC2 is the primary camera used for most of the Ceres observations and the basis of our work.  The camera has a pixel scale of 93.7 μrad, a 1024x1024 CCD detector, making a square field-of-view (FOV) of 5.5º on a side.  It is equipped with a wideband clear filter centered at 730 nm wavelength, and seven color filters centered at 439 nm to 965 nm with bandpasses of about 40 nm (about 90 nm for the 965 nm filter).

For the purpose of covering the whole surface of Ceres at the full spectral range of the FC, we used all color images collected during the first two science orbits: the "RC3" (rotational characterization 3) orbit at a radius of about 14,000 km and "Survey" orbit at a radius of about 4900 km.  Both orbits are circular polar orbits where the spacecraft moved from north pole towards south pole on the day side of Ceres, with the angle between the orbital plane and the Sun-Ceres line about 7º and 14º, respectively.  The RC3 observations included five observing sequences, two of which were executed on the night side of Ceres to search for dust near the surface of Ceres (Li et al., 2015), whereas the other three, termed RC3-equator, RC3-north, and RC3-south, were executed on the day side using all filters at the sub-spacecraft latitude near the equator and around 40º north and south, respectively.  We only included the RC3 images taken from May 4 to 7, 2015 on the day side of Ceres in our study.  Ceres filled about 70% of the FOV of FC2 in the RC3 images at a pixel scale of ~1.3 km/pixel.  In the Survey orbit, the FC captured images only on the day side using both clear and all seven color filters.  The FOV is about half the diameter of Ceres, and the pixel footprint is about 0.45 km.  The RC3 dayside and Survey images have higher spatial resolution in all color filters than those collected earlier during approach to Ceres.  Compared to those collected in later mission phases at lower altitude, the RC3 and Survey images cover a wide range of emission angles for the whole surface of Ceres with a minimal correlation between scattering angles and latitude, making a good set of data for a comprehensive study about the global photometric properties of Ceres.

The basic calibration of the FC2 images follows the steps outlined in Schröder et al. (2013b).  Images are calibrated to a dimensionless unit of radiance factor (RADF), which is the ratio between the brightness of a surface to that of a perfectly scattering Lambert surface of the same size and distance to the Sun and observer, but illuminated at normal direction (Hapke 1981).  RADF is synonymous to the commonly referred quantity *I/F*.  The FC color images are affected by an in-field stray light component (Schröder et al., 2014a; Kovacs et al., 2013), for which we did not make attempt to correct, but rather smoothed it out to some extent in the reduction of photometric data as will be discussed in detail in the next section.  All raw and calibrated data used in our study have been archived at Planetary Data System (PDS) Small Bodies Node (Nathues et al., 2015a; Nathues et al., 2016a).

*2.2. Photometric data reduction*

In order to fit the data to photometric models, which describe the dependence of RADF on scattering geometry ($i, e, \alpha$), we need to calculate the scattering geometry of all pixels in all images, extract the data and organize them in the form of RADF($i, e, \alpha$), and reduce in a way that best facilitates the model fitting of our purposes.

The local scattering geometry ($i, e, \alpha, \lambda, \phi$), with $\lambda$ and $\phi$ being geographic latitude and longitude, respectively, are calculated with the USGS Integrated Software for Imagers and Spectrometers ISIS3 (Anderson et al., 2004; Becker et al., 2012), which uses NAIF SPICE data archived at PDS (Krening et al., 2012) to determine the position and pointing of the spacecraft, the target, and the Sun. We used the shape model of Ceres derived primarily from the data acquired during Dawn's HAMO (high-altitude mapping orbit) phase of Dawn mission (Preusker et al., 2016; Roatsch et al., 2016a), which has a grid spacing of 0.135 km, or about 3× finer than the Survey data that we used in this photometric study, and covers about 98% of Ceres' surface with a vertical accuracy of about 10 m. The shape model is expressed in a Ceres-fixed reference frame that has the *z*-axis aligned with the rotational axis of Ceres and the prime meridian defined by the small crater Kait (Roatsch et al., 2016b).

Given the large number of images that we used, the photometric data from each filter contain about 42 million pixels in total, making it impractical to fit all together. Thus, we binned the data in ($i, e, \alpha$) space with a bin size of 5º for all angles, reducing the total number of data points to about 4000 in each filter. The photometric data points with $i>80º$ or $e>80º$ are discarded from the model fitting to avoid pixels too close to the limb or terminator. Schröder et al. (2017) demonstrated that 80º is a good cutoff for photometric data modeling that maximizes the surface coverage on Ceres, while still minimizing the registration uncertainty and the potential problem in photometric models near the limb and terminator. Fig. 1 shows the reduced photometric data from filter F2 at 555 nm effective wavelength as an example of the data that we fitted to models.

In order to map out the photometric model parameters across the surface of Ceres (Section 5), we divided the surface of Ceres into latitude-longitude grids of width 1º in size in both directions, and went through the geocentric coordinates of all pixels in all images and put the RADF($i, e, \alpha$) data into their corresponding grid. For each grid, we fitted a photometric model independently. Note that the changing physical area of grid with latitude does not affect photometric modeling results, although it will affect the number of data points in the grid and in turn the model quality. We did not project the images into latitude-longitude plane before extracting the photometric grid data as done by Schröder et al. (2013a, 2017) to avoid interpolation between pixels, although the effect of our averaging over the grid should be equivalent to interpolation.

The characteristics of the photometric grid data are shown in . In latitude between about ±50º, each 1º latitude-longitude grid contains more than 600 data points. The minimum incidence angles over the surface of Ceres have a strong correlation with latitude, which is unavoidable because incidence angle is determined by subsolar latitude that does not change much due to the low obliquity of Ceres (Russell et al., 2016). The maximum incidence angles are always greater than 80º, because the RC3 data always contain the whole surface of Ceres inside the FOV, thus covering the entire terminator. The coverage for emission angle is between a few degrees to >80º, again resulting from the full coverage of Ceres by the camera FOV in the RC3 data. For the distribution of the minimum phase angle, although some pattern is visible, the range is narrow with a width of about 3º. Because we do not plan to fit the opposition effect (see Sections 3 and 4), this distribution is not expected to have significant consequence on our modeling results. On the other hand, the maximum phase angle varies substantially across the surface, from <50º near the equator to nearly 90º towards the poles, with a strong latitudinal trend. The reason for this distribution and the latitudinal correlation is that only the RC3 data, which contains the whole Ceres disk in the FOV, can provide a uniform coverage in phase angle across the whole surface. However, the RC3 data were collected only near three discrete sub-spacecraft latitudes of 0º and ±40º, thus could reach a maximum phase angle of only <50º for the whole surface of Ceres. The Survey data, which provide coverage at higher phase angles when the spacecraft was at high latitude, are mostly nadir-pointed and contain only the center half of Ceres' disk in the FOV, missing the low-latitude region. Therefore, mid- to low-latitude regions do not have data at phase angles >50º. For this reason, we have to be cautious about the modeling related to the phase function, primarily the macroscopic roughness and SPPF, and check for any similar patterns between the resulting maps and the distribution of maximum phase angle to avoid interpreting modeling artifacts. Also, when we study the spatial variations of parameters, we should compare locations at similar latitudes.

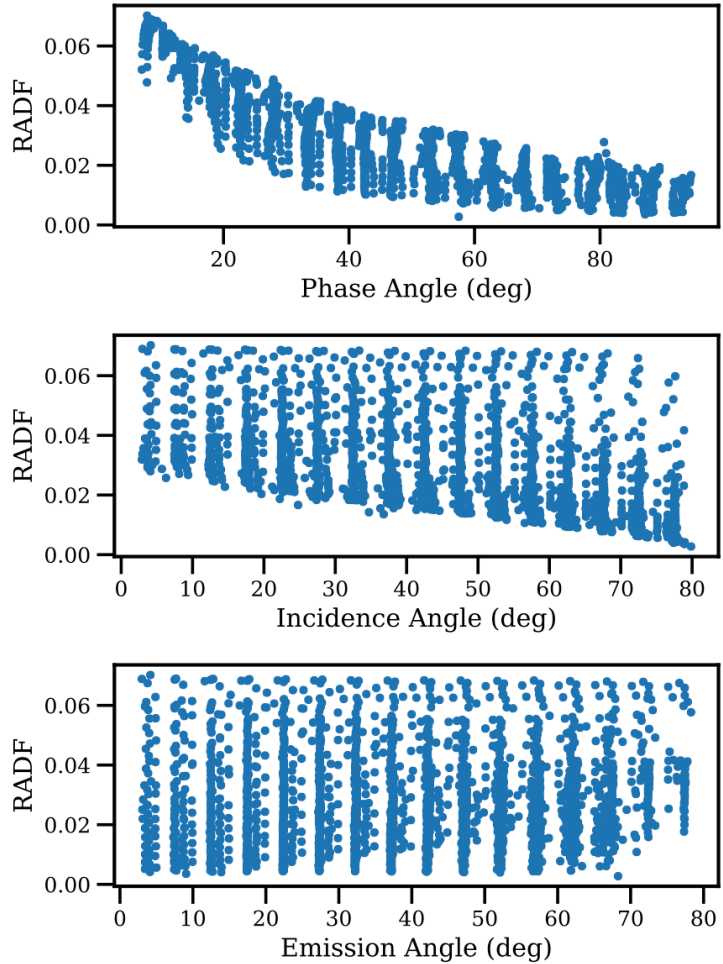

Figure 1. Reduced photometric data from filter F2. The three panels show RADF plotted with respect to phase angle (upper), incidence angle (middle), and emission angle (lower). Data points with $i>80º$ or $e>80º$ are discarded.

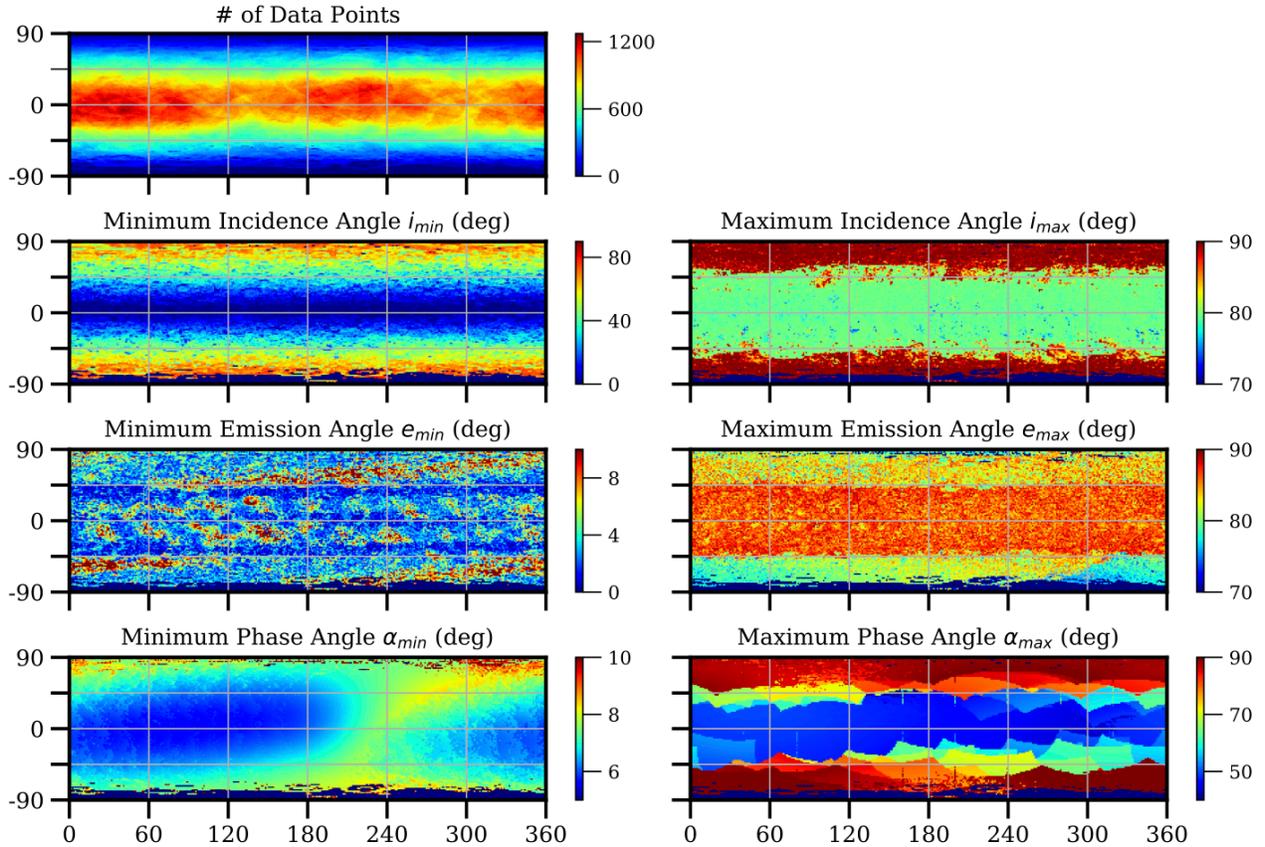

Figure 2. Photometric grid data characteristics. Panel content is noted on the top of every panel. Note that their color bar scales are all different.

The characteristics of stray light have been analyzed by Schröder et al. (2014a) and Kovacs et al. (2013). Stray light increases the scene brightness by up to 10-14% for filters F4 (917 nm), F6 (829 nm), F7 (653 nm), and F8 (438 nm), and up to 4-6% for the other three filters. The spatial distribution of stray light in the FOV depends on the brightness distribution of the scene and is not uniform in RC3 and Survey images, especially those containing limb and/or terminator. Therefore, stray light could affect photometric modeling in two aspects:
1. It increases the modeled albedos by increasing the scene brightness;
2. It changes the distribution of brightness with respect to scattering geometry.

On the other hand, the photometric data reduction process as described above effectively averages all the pixels that are within the same scattering geometry bin but could distribute all over the FOVs from many images. Therefore, the different effects of stray light in the RADF($i$, $e$, $\alpha$) data from different images should be smoothed out to some extent, and the net results are an increased model albedo than the true value by roughly the fraction of stray light, and an increased model scatter. Given that reflectance is proportional to albedo for a dark surface like Ceres', we just need to scale our modeled albedo based on the estimate of stray light contributions for respective filters (Schröder et al., 2014a) to derive the true albedo. Other parameters that describe the ($i$, $e$, $\alpha$) dependence of RADF should not be affected, including the phase function, because the measured RADF is increased by stray light by the same scaling factor at different scattering geometries, equivalently an effect of increased albedo. In our discussions of the modeling results, we will avoid basing our analysis on the absolute values of the best-fit parameters unless they are consistent with previous modeling values, in order to minimize the impact of stray light on our conclusions.

## 3. Photometric models

Schröder et al. (2017) have demonstrated that, among the photometric models that they tested, the Hapke model and the Akimov model are the best to describe the photometric behaviors of Ceres. Therefore, we base our analysis primarily on the framework of the Hapke model, as well as the Akimov disk-function coupled with a linear magnitude phase function in our photometric model mapping. We also include the Lommel-Seeliger (LS) disk-function in our analysis for its simplicity.

### 3.1. Hapke model

We adopted a form of Hapke model as follows,

$$RADF(i, e, \alpha) = \frac{w}{4} \frac{\mu_{0e}}{\mu_{0e}+\mu_e} [B_{SH}(B_0, h; \alpha)p(\alpha) + H(\mu_{0e}, w)H(\mu_e, w) - 1]S(\bar{\theta}; i, e, \alpha) \quad \ldots(1)$$

In this form, $\mu_{0e}$ and $\mu_e$ are the cosines of local $i$ and $e$ corrected for roughness, $\bar{\theta}$, respectively. $B_{SH}$ is the shadow-hiding opposition effect with two parameters, the amplitude, $B_0$, and width, $h$. The form of $B_{SH}$ adopted here is the same as previously used in Li et al. (2004; 2006). $H(\mu, w)$ is the Chandrasekhar H-function, where $H(\mu_{0e}, w)H(\mu_e, w) - 1$ characterizes multiple scattering assuming isotropic single-scattering. We adopted the approximated form of H-function suggested by Hapke (2002). $S(\bar{\theta}; i, e, \alpha)$ is the correction for surface roughness, $\bar{\theta}$. We followed the formalism of roughness correction as in Hapke (1984). $p(\alpha)$ is the SPPF, which could take a 1pHG form that has a single parameter called asymmetry factor, $\xi$,

$$p(\xi; \alpha) = \frac{1-\xi^2}{(1+2\xi \cos \alpha + \xi^2)^{3/2}} \quad \ldots(2)$$

where $-1 \leq \xi \leq 1$, characterizing the spatial distribution of the scattered light from a single particle with respect to 90º phase angle, with $\xi < 0$ associated with predominantly backscattering, $\xi > 0$ associated with predominantly forward scattering, and $\xi = 0$ isotropic scattering. When the SPPF takes this form, the Hapke model as in Eq. (1) has a total of five parameters. Alternatively, the SPPF could take a 2pHG form with two parameters, $b$ and $c$,

$$p(b, c; \alpha) = \frac{1+c}{2} \frac{1-b^2}{(1-2b \cos \alpha + b^2)^{3/2}} + \frac{1-c}{2} \frac{1-b^2}{(1+2b \cos \alpha + b^2)^{3/2}} \quad \ldots(3)$$

where $0 \leq b \leq 1$ and $-1 \leq c \leq 1$. The first term represents backward scattering, while the second term represents forward scattering. Parameter $b$ determines the strength of the anisotropy of the phase function, with larger values indicating stronger anisotropy; whereas parameter $c$ determines whether the scattering is predominantly backward ($c > 0$) or forward ($c < 0$), or symmetric ($c = 0$). The asymmetry factor, $\xi = -bc$, has the same meaning as for 1pHG. This form of $p(\alpha)$ makes the Hapke model have six parameters in total. Note that the $c$ parameter here needs to be linearly scaled to range [0, 1] in order to be consistent with the 2pHG in the Hapke model form adopted by the USGS ISIS3 software. In our modeling effort, we tried both 1pHG and 2pHG SPPF for the purposes of consistency check and better understanding the photometric behaviors of Ceres.

Hapke (2002) updated the model by considering anisotropic multiple scattering. For a dark surface with a geometric albedo of about 0.10 (Li et al., 2016b; Schröder et al., 2017), we expected multiple scattering to play a minor role, and decided not to include anisotropic multiple scattering in our modeling. Hapke (2002) also added coherent backscattering opposition effect (CBOE) to the model. CBOE generally appears at phase angles <2º, while our data, with a minimum phase angle of about 7º, do not allow the determination of CBOE. In addition, CBOE is a multiple scattering phenomenon, which is expected to be weak on a dark surface like Ceres'. Therefore, we did not include CBOE in our model. Hapke (2008) further considered the effect of porosity in the optically active regolith. We did not include porosity in our modeling effort because for a dark surface, the porosity parameter is equivalently a scaling factor for the reflectance and cannot be separated from SSA, and because the lack of data within the opposition geometry prevents us from deriving the porosity.

*3.2. Empirical model*

In the simple type of empirical models, reflectance RADF is separated into two parts, the equigonal albedo and the disk function (Kaasalainen et al., 2001; Shkuratov et al., 2011),

$$RADF(i, e, \alpha) = A_{eq}(\alpha) D(i, e, \alpha) \quad \ldots(4)$$

where the disk-function, $D(i, e, \alpha)$, describes the dependence of RADF on local topography ($i$, $e$), which could depend on $\alpha$. In our analysis, the disk-function takes either the LS function model,

$$D(i, e) = 2 \frac{\cos i}{\cos i + \cos e} \quad \ldots(5)$$

or the parameter-less Akimov disk function model (Shkuratov et al., 2011),

$$D(\alpha, b, l) = \cos \frac{\alpha}{2} \cos \left[ \frac{\pi}{\pi - \alpha} \left( l - \frac{\alpha}{2} \right) \right] \frac{(\cos b)^{\alpha/(\pi - \alpha)}}{\cos l} \quad \ldots(6)$$

where $b$ and $l$ are photometric latitude and longitude, respectively. Same as the LS function, the Akimov model results in a disk of constant brightness at $\alpha = 0º$, or equivalently the same values for normal albedo and geometric albedo.

After correcting for the disk function, the equigonal albedo $A_{eq}$ depends only on phase angle. We adopted a linear model in magnitude space to describe $A_{eq}(\alpha)$,

$$A_{eq}(\alpha) = A_n 10^{-0.4\beta\alpha} \quad \ldots(7)$$

where $A_n$ is normal albedo, and $\beta$ is the phase slope parameter in mag/deg.

We note that this linear-magnitude phase function (Eq. 7) is essentially an exponential phase function model, same as the one adopted by Schröder et al. (2017) but with a different scaling factor from their slope parameter, $\nu$, and the modeling results can be directly related by $\nu = 52.77\beta$. In our photometric model mapping (Section 5), we included both $\beta$ and $\nu$ parameters to compare with the previous results in order to confirm the features that we observed. We did not apply these empirical models for global photometric modeling (Section 4).

*3.3. Model fitting*

The best-fit photometric model is defined in a $\chi^2$ sense. We defined the relative root mean square (RMS) to quantify the model quality,

$$Rel.RMS = \frac{1}{\bar{r}}\sqrt{\frac{1}{n}\sum_{i=1}^{n}(r_i - r_{i,model})^2} \qquad \ldots(8)$$

where $r_i$ is the measured RADF, and $r_{i,\,model}$ is the modeled RADF, the sum is over all $n$ data points, and $\bar{r}$ is the average RADF of all data points. The minimization of RMS is performed with the Levenberg-Marquardt algorithm with constrained search space for the model parameters (Moré, 1978; Markwardt, 2009).

Because of the inter-correlation between the Hapke parameters, sometimes the fit converges to a local minimum rather than the global minimum. To avoid this potential problem, we performed our model fitting with at least 100 trials with randomly generated initial parameters. For more than 90% of the trials, the models were able to converge to a small area around the best-fit model. We used the same curve fitting algorithm for empirical models as for the Hapke model.

## 4. Global photometric modeling

We focused on the Hapke models to derive the global photometric properties of Ceres. The minimum phase angle of about 7º in our data does not allow us to reliably model the opposition effect. Even with the data within the opposition acquired in April 2017, the Hapke modeling still could not return a satisfactory fit with reasonable opposition effect parameters either (Schröder et al., in press). Therefore, we tried two cases in the model fitting: 1) fixing $B_0$=1.6 and $h$=0.06 as found by Helfenstein and Veverka (1989), and 2) set free both parameters. We also fitted the data with both 1pHG and 2pHG SPPF in the Hapke model, making a total of four cases to compare. The best-fit parameters of all seven color filters are plotted in Fig. 3.

As indicated by the RMS, the models with 2pHG perform consistently better than those with 1pHG. Inspecting the ratio of measured RADF to modeled RADF with respect to scattering angle reveals an obvious trend with phase angle for the 1pHG model (Fig. 4a), but not for the 2pHG Hapke model (Fig. 4c). In either model form, the ones with free opposition parameters performed better than the ones with fixed parameters, simply because of more degrees of freedom allowed in the former. For both model forms, when the opposition parameters were set free, the $B_0$ parameters always ended up at the imposed upper limit of 6.0 (Fig. 3). On the other hand, the model quality of those two cases for the 2pHG Hapke model is close to one another. These observations suggest: 1) 2pHG was necessary to model the photometric behavior of Ceres, even though the maximum phase angle in the data that we fitted was just about 95º. ; 2) $B_0$ and $h$ could not be constrained from our data; 3) Because the photometric parameters in Hapke model are entangled, perhaps except for $\bar{\theta}$, which is mostly determined by the ($i$, $e$) dependence of reflectance and thus to a less extent entangled with others, caution has to be used when comparing the photometric parameters at different wavelengths and with other objects.

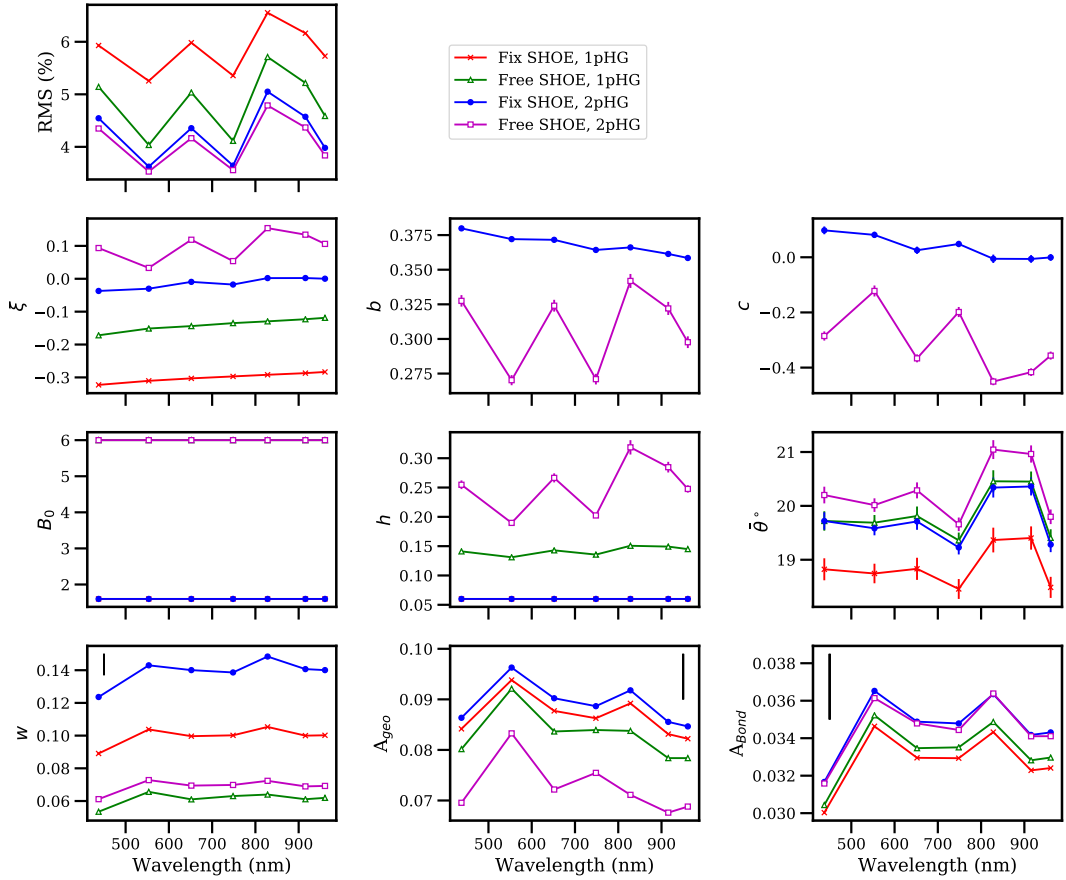

Figure 3. Best-fit Hapke parameters for all four model cases (see text). The SSA plot is corrected for stray light by a simple scaling as described in Section 2.2, although the bumps at 555 nm and 830 nm suggest that the correction may not be clean. The plots of *b* and *c* are for 2pHG model only. The fits with the opposition parameter $B_0$ and *h* free all result in $B_0$=6.0, which is the upper limit imposed in the model fitting, and the two lines are on top of one another. The statistical error bars from the model fit itself are plotted, but in most cases are smaller than the symbols and not visible. The vertical bars in the three plots for SSA, $A_{geo}$, and $A_{Bond}$ (the three plots in the bottom row) represent the approximate photometric calibration error bars of 5%. See text for a full analysis of the modeling uncertainties.

Because the 2pHG case with fixed opposition parameters has similar quality as the case that allows the opposition parameters to change, and because the latter results in very noisy parameter spectra for *h* and *ξ*, we decided to base our analysis of the modeling results primarily on the results from 2pHG Hapke model with fixed opposition parameters (Table 1, filled blue circles in Fig. 3). The model parameters for 1pHG Hapke model with fixed $B_0$ and *h* parameters are also reported in Table 2 for the purpose of comparing with previous Hapke model analyses of other asteroids, almost all of which have been performed with the 1pHG form.

Table 1. Best-fit parameters of Ceres with the Hapke model using a 2pHG. $B_0$=1.6 and $h$=0.06 are fixed in the model fitting. The albedos listed here are all corrected for stray light by scaling (see Section 2.2 for details). The effective wavelengths of filters are based on Schröder et al. (2013b). See Section 4.1 for the discussion of uncertainty estimates.

| Filter | $\lambda$ (nm) | $w$ | $b$ | $c$ | $\xi$ | $\bar{\theta}$ (º) | $A_{geo}$ | $A_{Bond}$ | RMS (%) |
|---|---|---|---|---|---|---|---|---|---|
| F2 | 555 | 0.143 | 0.372 | 0.081 | -0.030 | 19.6 | 0.096 | 0.037 | 3.6 |
| F3 | 749 | 0.139 | 0.364 | 0.048 | -0.018 | 19.2 | 0.089 | 0.036 | 3.6 |
| F4 | 917 | 0.141 | 0.361 | -0.006 | -0.002 | 20.4 | 0.086 | 0.034 | 4.6 |
| F5 | 965 | 0.140 | 0.358 | -0.001 | 0.000 | 19.3 | 0.085 | 0.034 | 4.0 |
| F6 | 829 | 0.148 | 0.366 | -0.006 | 0.002 | 20.3 | 0.092 | 0.036 | 5.1 |
| F7 | 653 | 0.140 | 0.372 | 0.025 | -0.009 | 19.7 | 0.090 | 0.036 | 4.4 |
| F8 | 438 | 0.124 | 0.380 | 0.098 | -0.037 | 19.7 | 0.086 | 0.032 | 4.5 |
| Error | | -0.04 +0.05 | ±0.06 | -0.08 +0.05 | | ±6 | ±0.005 | ±0.002 | |

Table 2. Best-fit parameters of Ceres with the Hapke model using 1pHG. $B_0$=1.6 and $h$=0.06 are fixed in the model fitting. The albedos listed here are all corrected for stray light by scaling (see Section 2.2 for details). Th effective wavelengths of filters are based on Schröder et al. (2013b). See Section 4.1 for the discussion of uncertainty estimates.

| Filter | $\lambda$ (nm) | $w$ | $\xi$ | $\bar{\theta}$ (º) | $A_{geo}$ | $A_{Bond}$ | RMS (%) |
|---|---|---|---|---|---|---|---|
| F2 | 555 | 0.104 | -0.310 | 18.7 | 0.094 | 0.035 | 5.3 |
| F3 | 749 | 0.100 | -0.297 | 18.5 | 0.086 | 0.033 | 5.4 |
| F4 | 917 | 0.100 | -0.287 | 19.4 | 0.083 | 0.032 | 6.2 |
| F5 | 965 | 0.100 | -0.283 | 18.5 | 0.082 | 0.032 | 5.7 |
| F6 | 829 | 0.105 | -0.292 | 19.4 | 0.089 | 0.034 | 6.6 |
| F7 | 653 | 0.100 | -0.303 | 18.8 | 0.088 | 0.033 | 6.0 |
| F8 | 438 | 0.089 | -0.323 | 18.8 | 0.084 | 0.030 | 5.9 |
| Error | | -0.04 +0.05 | ±0.05 | ±6 | ±0.005 | ±0.002 | |

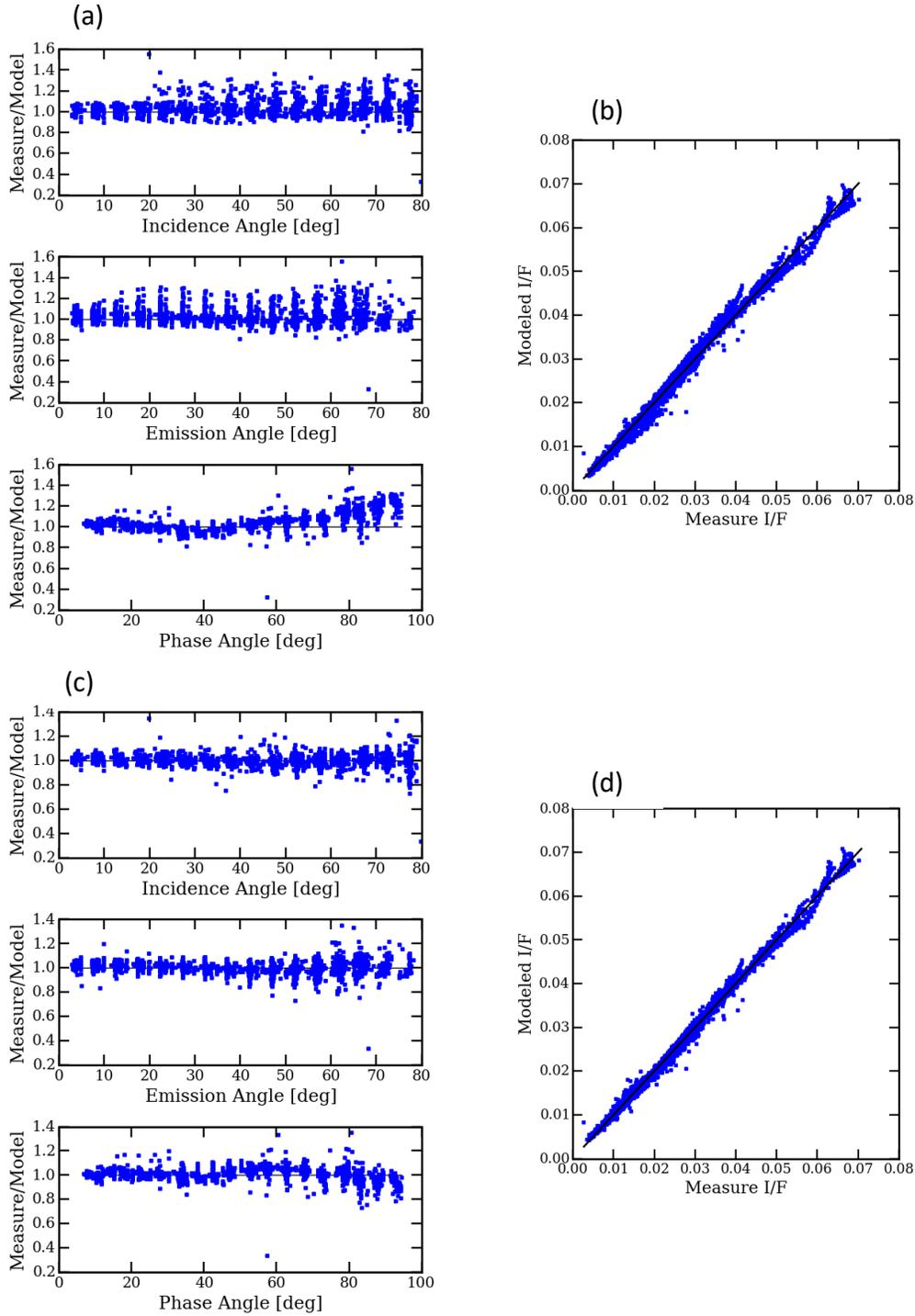

Figure 4. Quality plots of the Hapke model fitting with the F2 filter data (555 nm). Panels (a) and (b) are for 1pHG model, and panels (c) and (d) are for 2pHG model. In these two cases, the opposition parameters are fixed at $B_0$=1.6 and $h$=0.06. The ratio between measured RADF and modeled RADF with respect to phase angle $\alpha$ for the 1pHG (panel a) shows an obvious systematic trend, which does not appear in the plot for 2pHG (panel c). There is no systematic trend with respect to $i$ and $e$ for either the 1pHG or 2pHG form of the Hapke model. The model RMS's are 5.3% and 3.6% for the 1pHG and 2pHG cases, respectively.

*4.1. Model uncertainty*

Because of the complicated entanglement among the Hapke parameters, their model uncertainties cannot be directly derived from statistical principles of least-$\chi^2$ fit. We estimated the uncertainties following the similar approach by Helfenstein and Shephard (2011) and Li et al. (2013). We fixed the value of the parameter under consideration in a range surrounding the best-fit value, and fitted the remaining parameters (still with $B_0$ and $h$ fixed) to find the $\chi^2$'s, which is essentially the term inside the square root in Eq. 8. Then the 1-$\sigma$ uncertainty range for this particular parameter is defined as the locus where $\chi^2$ is less than twice the minimum $\chi^2$. An example for the uncertainty estimate is shown in Fig. 5 for the roughness parameter.

In addition, we visually inspected how the model fitting worsens when perturbing the parameter under consideration away from the best-fit value, to judge whether the uncertainty estimates are sensible. Different parameters have to be inspected with different approaches. For the roughness parameter, we compared the model fitting to the brightness scans along photometric equators and mirror meridians at various phase angles, similar to the experiment in Li et al. (2013). For the phase function parameters, $b$ and $c$, and for the SSA, we compared the data, after correcting for the LS term $\mu_{0e}/(\mu_{0e} + \mu_e)$ and roughness correction $S(\bar{\theta}; i, e, \alpha)$, with the surface phase function model $B(B_0, h; \alpha)p(\alpha) + H(w, \mu_{0e})H(w, \mu_e) - 1$. The inspection suggests that our error estimates are reasonable.

The formal uncertainties that we derived are similar for all bands: about ±6º for the roughness parameter; about ±0.06 for the phase function parameter $b$; about -0.08 and +0.05 for parameter $c$, asymmetric with respect to the best-fit values; and about -0.04 and +0.05 for the SSA. Note that we should consider these error bars systematic in the sense that they do not represent the relative model scatter from one band to the next. The error estimate that we discussed here is related to how well the model describes the photometric behavior of Ceres' surface, given the measurement noise. For the case of SSA, this error estimate does not include the absolute flux calibration of the camera, which is about 5% (Schröder et al. 2013b), nor the uncertainty as introduced by stray light, which is estimated to be about 1-2% based on Schröder et al. (2014a). The uncertainty of geometric albedo is about 6% considering all the sources of errors, and the uncertainty of Bond albedo is dominated by the absolute calibration uncertainty to be 5%. On the other hand, the scatter in the spectrum of the best-fit parameter is a good measurement of the robustness of the wavelength trend. Therefore, although the systematic errors are all much larger than the ranges of variations in the spectra for the best-fit parameters, as long as the scatter is small enough compared to the overall wavelength trend, we consider that such trend reflects the real wavelength dependence of Ceres' photometric behavior.

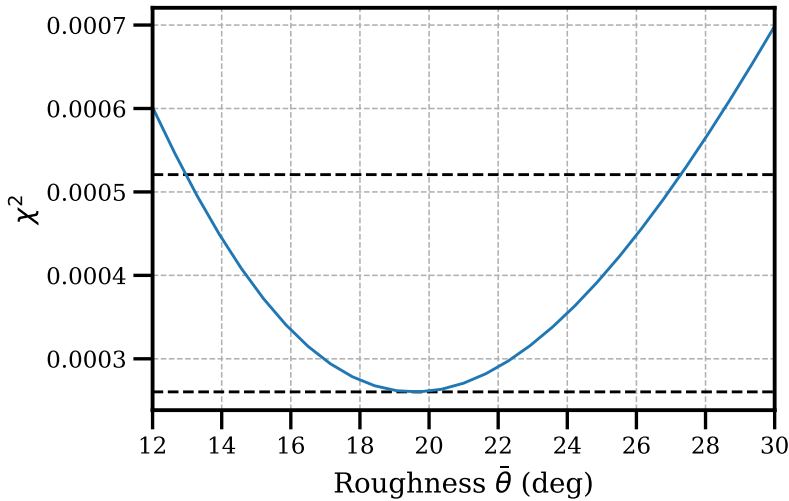

Figure 5. $\chi^2$ plot with respect to fixed roughness parameter $\bar{\theta}$ as an example for our uncertainty estimate of Hapke model parameters. The lower and upper horizontal dashed lines mark the position of minimum $\chi^2$ and of twice of the minimum. The range of uncertainty for $\bar{\theta}$ is estimated to range from 13º to 27º.

*4.2. Phase Function*

As shown in Fig. 6a, compared to 1pHG, the best-fit 2pHG function for Ceres results in a disk-integrated phase function that decreases more steeply at moderate phase angles from 20º to 60º, then curves up at higher phase angles. This behavior is also evident in the systematic trend of the ratio between measured RADF and modeled RADF with respect to phase angle, where when using 1pHG to fit the data, the measurement is lower than the best-fit model at moderate phase angles while higher at higher phase angles (Fig. 4a). The use of 2pHG removed such a systematic trend (Fig. 4c), and resulted in a lower RMS that is statistically significant. Therefore, we conclude that the phase function of Ceres can only be satisfactorily characterized by the 2pHG but not the 1pHG.

For both the 1pHG and 2pHG modeling, the disk-integrated phase function of Ceres shows dependence on wavelength where the strength of backscattering decreases with wavelength monotonically from 438 nm to 965 nm (Fig. 6b, c, d). This wavelength trend is consistent with phase reddening, which for Ceres was first reported by Tedesco et al. (1983) from ground-based data. Li et al. (2016b), based on the measurements from all the previous ground-based data that they could find, showed that the spectral slope of Ceres monotonically increases with phase angle to at least 20º phase angle. Most recently Ciarniello et al. (2017), Longobardo et al. (2018) also reported phase reddening of Ceres based on Dawn data.

While the existence of stray light prevents us from quantifying phase reddening of Ceres and comparing it with other objects, we can still qualitatively characterize it based on the wavelength dependence of the phase function, because there is no monotonic wavelength dependence for stray light (Schröder et al., 2014a). First, the monotonic decrease of the phase slope of Ceres with wavelength is different from that of Vesta, whose phase slope decreases until 750 nm, which is just outside of its 1-µm mafic band where its spectrum starts to turn down, then increases towards 965 nm, which is near the center of the 1-µm band (Li et al., 2013). The phase reddening on Vesta appears to depend on its spectral slope, where positive spectral slope corresponds to phase reddening and negative spectral slope corresponds to phase bluing. While for Ceres, the spectrum is flat across the wavelength range of our data (cf. Rivkin et al., 2011; Nathues et al., 2015b), yet the strength of phase reddening seems to be comparable to or even slightly stronger than that of Vesta as judged from the phase function ratio plot (Fig. 6c, d). This difference suggests that albedo is not a dominant cause of phase reddening for Ceres. We will further discuss this phenomenon in Section 6.3. Second, the phase function ratio curves of Ceres have different shapes from those of Vesta. The indications are that at phase angles lower than 20º, which is approximately the maximum phase angle accessible from the ground, Vesta displays stronger phase reddening than Ceres. This is consistent with observations (Reddy et al., 2011; Li et al., 2016b). On the other hand, at higher phase angles, especially >80º, Ceres could have stronger phase reddening than Vesta. This result can be tested with Dawn VIR data of both objects taken at high phase angle.

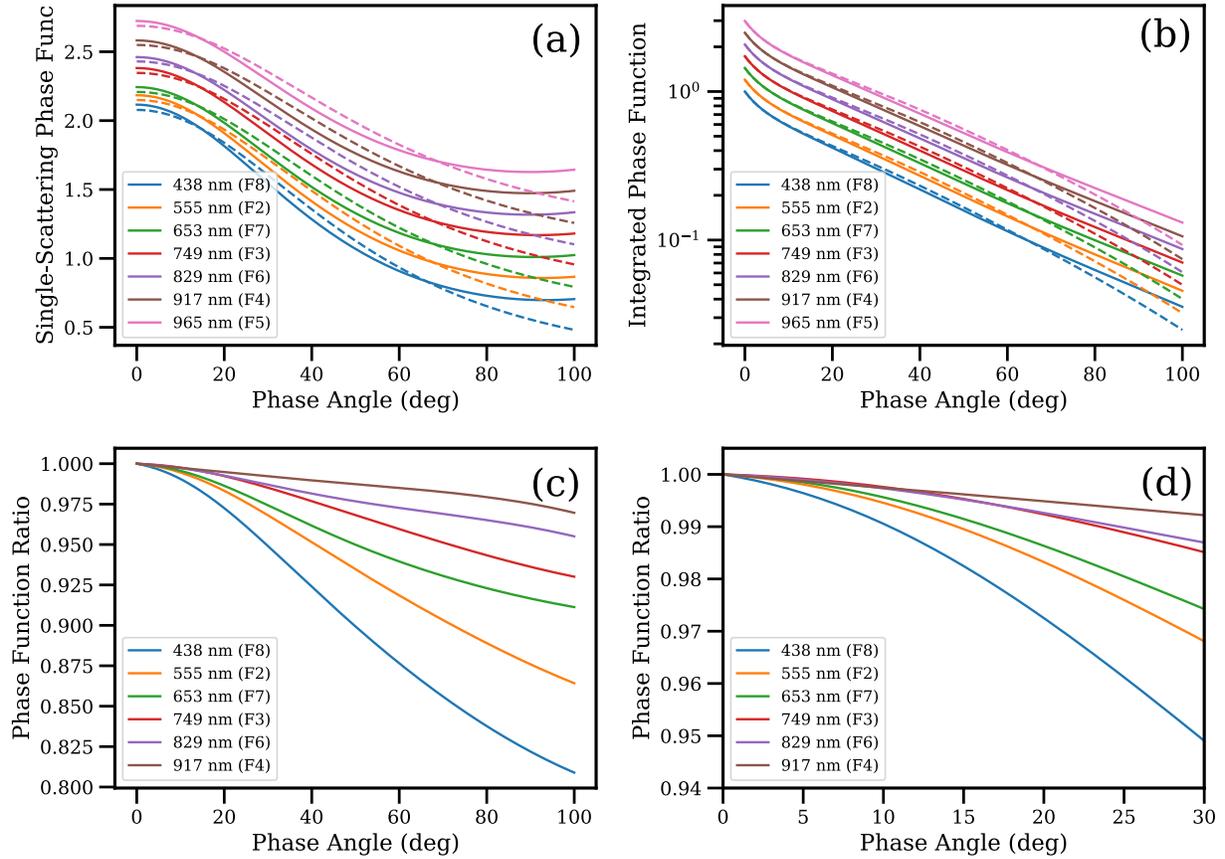

Figure 6. Panel (a) is the best-fit single-particle phase function to Ceres data in all seven color filters. Solid lines are the 2pHG, and dashed lines are the 1pHG scaled by the ratio of the SSA fitted with the 1pHG to that fitted with the 2pHG. The lines of 438 nm are plotted at the original *y*-scale, while the lines for all other bands are shifted upward by an increment of 0.1 in *y*-axis for clarity. Panel (*b*) is the corresponding disk-integrated phase function, with the same legend as panel (a). All phase functions are normalized to unity at opposition, with the *y*-scale of the phase curves of 438 nm at the original scale and all other lines scaled upward by an increment of 20% in *y*-axis for clarity. Panel (c) is the ratio of disk-integrated phase function to the one at 965 nm (the longest wavelength in our dataset). Panel (d) is the same as panel (c) but zoomed in to show phase angles between 0º and 30º.

*4.3. Roughness*

Surface roughness affects the photometric behavior of a surface in two aspects: it changes the dependence of reflectance on local topography ($i$, $e$), and it decreases the forward scattered light, i.e., increases the slope of the surface phase function. The effects of roughness increase with phase angle, thereby a reliable determination of roughness requires disk-resolved data at moderate to high phase angles, preferably > 60º (Helfenstein, 1988, Helfenstein et al., 1988). As a geometric parameter, roughness itself should be independent of wavelengths. For a very bright surface where multiple scattering substantially diminishes shadows, the modeled value of roughness could be lower than true value. In this case, if the surface has a strongly sloped spectrum, then the modeled roughness could show a wavelength dependence. Neither case applies to Ceres.

In our modeling, the roughness parameter is consistently modeled to be within a narrow range of 18º to 21º without significant wavelength dependence, consistent with it being a geometric parameter. The average roughness of 20º±6º is consistent with the values previously derived based on Dawn data (Li et al. 2016a, Schröder et al. 2017, Ciarniello et al. 2017). A very high value of 44º±5º was reported by Li et al. (2006) based on HST data. However, that value could be unreliable for two reasons: 1. The HST data were taken at low phase angles between 5º and 8º, where the effect of roughness is weak; and 2. The camera that they used, the High-Resolution Channel of the Advanced Camera for Surveys, has a wide point-spread-function (PSF) that encircles <80% energy even in a 10 pixel radius aperture (Avila et al., 2017). Such a PSF results in a significant limb darkening for the extended disk of Ceres, which was just about 30 pixels in diameter in those HST images.

*4.4. Albedo*

All modeled albedo quantities, including the SSA, geometric albedo, and Bond albedo are strongly dependent on the photometric calibration of the data. As mentioned before, stray-light affects the photometric calibration of FC images. Even though we tried to account for it by a simple scaling based on Schröder et al. (2014a) in our modeled albedo quantities, the effect is still evident from the scatter of points in the albedo spectra ( ). Despite the scatter, the overall shapes of all albedo spectra are consistent with ground-based observations, and the blue slope of the geometric albedo spectrum is consistent with previous results (Li et al., 2016).

The SSA of Ceres is 0.14±0.04 at 555 nm, based on the 2pHG model with fixed opposition parameters. This value has an excellent agreement with that derived from the VIR data (Ciarniello et al., 2017), which used exactly the same form of Hapke model as we did. On the other hand, this value of SSA is much higher than previous modeling results from ground (Reddy et al., 2015) and HST data (Li et al., 2006). We suspect that such a difference is caused by the use of 1pHG in their modeling. In our modeling attempts with the 1pHG, the derived SSA was closer to the previously derived values, although still higher (Table 2). With data covering a much wider range of phase angles than before and a 2pHG that appears to systematically better fit the data than a 1pHG, we consider the value we derived here more reliable than previous modeling results. The geometric albedo of Ceres based on the best-fit Hapke parameters is 0.096±0.006 at 555 nm, which is consistent with previous determinations (Reddy et al., 2015, Li et al., 2006, Ciarniello et al. 2017), and in an excellent agreement with the measurement from opposition (0.094±0.005; Schröder et al., in press). We note that the modeled geometric albedo here is based on an assumed opposition effect, and the agreement is a coincidence to some extent. On the other hand, the Bond albedo depends on the overall shape of the phase function, and thus can generally be more reliably determined than geometric albedo, as indicated by the consistent results from all modeling cases (Fig. 3). The Bond albedo of Ceres at 555 nm is 0.037±0.002, and the uncertainty is completely dominated by the calibration uncertainty of the FC data. Given the flat spectrum of Ceres across visible and near-infrared, we can use this value as its bolometric Bond albedo, too.

## 5. Photometric model mapping

The traditional approach of studying the photometric variations on the surface of an object is through "photometric mapping", that is, to fit a photometric model for the whole area of interest, then use that model to correct images to a common viewing and illumination geometry, and finally mosaic images together to generate a reflectance map of the area. This approach implicitly assumes that all photometric properties other than albedo are uniform, or, equivalently, it folds the variations in all other photometric properties into the equivalent variations of albedo (Li et al., 2015).

With sufficient data available, it is possible to study the variations in photometric properties other than albedo. As the first attempt of this kind for solar system small bodies, Li et al. (2007) fitted the Hapke model to individual terrains on comet 19P/Borrelly and reported large variations in albedo, phase function, and roughness, although their mapping may not be reliable given the small amount of images available from flyby observations and the small size of the terrains that they defined relative to the image resolution. Schröder et al. (2013a) and Schröder et al (2017), using Dawn observations of Vesta and Ceres, respectively, fitted an exponential phase function model (Eq. 7) to the photometric data for each latitude-longitude grid after corrected for the dependence on ($i$, $e$) with the Akimov disk-function, and derived the maps of both normal albedo and phase slope. Longobardo et al. (2014) also discriminated Vesta areas characterized by a different phase slope. The successful mapping process allowed them to analyze the maps in the context of geology and geomorphology for both objects.

The simple exponential model adopted by Schröder et al. (2013a, 2017) and the empirical approach adopted by Longobardo et al. (2014) cannot distinguish between the effects of surface roughness and the particle phase function, because both would change the slope of phase function in a similar manner and the model uses one single parameter to describe the phase slope. In addition, because roughness could change the disk-function of a surface, the use of a parameter-less disk-function such as the LS model or the parameter-less Akimov model could miss such effects. In this work, we pursued a similar mapping process but with the more sophisticated Hapke model, with the hope of separating the variations due to roughness and particle phase function. We refer to this process as "photometric model mapping" to distinguish it from the traditional approach of "photometric mapping".

On the other hand, caution has to be used when interpreting the maps of Hapke parameters. While it is generally accepted that the Hapke model is able to describe the general scattering behaviors of particulate surfaces, the true physical meanings of the model parameters have always been under intensive investigation and debate (e.g., Shepard and Helfenstein, 2007; 2011; Shkuratov et al., 2012; Hapke, 2013; etc.). For example, although the roughness parameter affects the disk-function and improves the fit to reflectance data with respect to local topography, it is never entirely clear what its true physical indications to planetary surfaces are and at what size scale (Helfenstein, 1988; Shepard and Campbell, 1998; Helfenstein and Shepard, 1999). In some work the roughness parameter has been dropped entirely, and its effect on phase function has been included in the phase function parameters (e.g., Shepard and Helfenstein, 2011). Another example is the SPPF, which Shkuratov et al. (2012) criticized as non-physical because of the truncation of the Fraunhofer diffraction peak in the forward scattering direction in the Hapke model treatment, whereas Hapke (2013) refuted that such peak is altered when isolated particles are brought near or into mutual contact with other particles in a regolith surface. Given these debates, we shall be careful about the interpretations of the parameter maps, and always refer to the geological and geomorphological context as well as the laboratory results. In particular, we consider that the roughness parameter is introduced as a separate parameter because it has an effect on the disk-function that cannot be fully compensated by any other parameters. Variations in this parameter should indicate variations of one or some physical properties, even though the particular mechanism is unclear. Our interpretations of SPPF will also be mostly based on relevant laboratory studies (e.g., McGuire and Hapke, 1995; Souchon et al., 2011; Pommerol et al., 2013; Pilorget et al., 2016).

In order to assess the robustness of this mapping process, we considered four models: 1) the LS disk-function (Eq. 5) and the linear phase function in magnitude (Eq. 7); 2) the Akimov disk-function (Eq. 6) and the linear phase function in magnitude (Eq. 7); 3) the Hapke model using 1pHG (Eqs. 1 and 2); and 4) the Hapke model using 2pHG (Eqs. 1 and 3). With much fewer data points in each latitude-longitude grid than the global photometric modeling, we had to limit the data in each grid to $i<60º$ and $e<60º$ in order to better avoid extreme geometries to ensure the model fitting quality. Modeling with a cutoff at 80º results in nearly twice as high relative RMS and noisy parameter maps that are hard to interpret. The fitting yields a number of maps for every case: the relative RMS map, the maps of all parameters of the corresponding model, and the normal, geometric, and Bond albedo maps. With the model parameter maps produced for all seven FC color filters, we were also able to study the spatial variations of the spectrum of every photometric parameter. Note, however, that the extremely bright Cerealia Facula inside Occator crater is saturated in many of the images we used, and therefore the modeling for that feature is not reliable. We do not include this feature in our discussion in this article. In addition, in our analysis of the photometric parameter maps, we focus on the global surface of Ceres and features larger than tens of km in size due to the 1º resolution in our latitude-longitude grid, which corresponds to 8 km near the equator.

*5.1. Mapping with empirical models*

Before applying photometric model mapping with the Hapke model, we performed mapping with the Akimov disk-function (Eq 6) and the LS disk-function (Eq. 5), coupled with a simple linear magnitude phase function model (Eq. 7). The resulting maps with Akimov disk-function model are displayed in Fig. 7. The relative RMS are generally between 2-5%, and for the band between ±40º latitude <3%, indicating good model fitting. The normal albedo map and phase slope map are entirely consistent with those derived by Schröder et al. (2017) with the same modeling process but using RC3 data only. With this sanity check, we are confident that our photometric model mapping process was able to produce results as expected.

The mapping results using the LS disk-function are similar to the Akimov model mapping results, with only slight differences (Fig. 8). The largest difference in the normal albedo map appears in the ejecta field to the northwest side of Occator crater, where the LS model results in a slightly lower albedo. The overall absolute scales of normal albedo maps are similar. The phase slope derived from the LS model is overall higher (steeper phase slope) than that derived from the Akimov model by about 10%. The model RMS map is slightly higher than that of the Akimov model map by about 1%. The higher model RMS is also consistent with the remark by Schröder et al. (2017) that the Akimov disk function performs better than the LS function for Ceres.

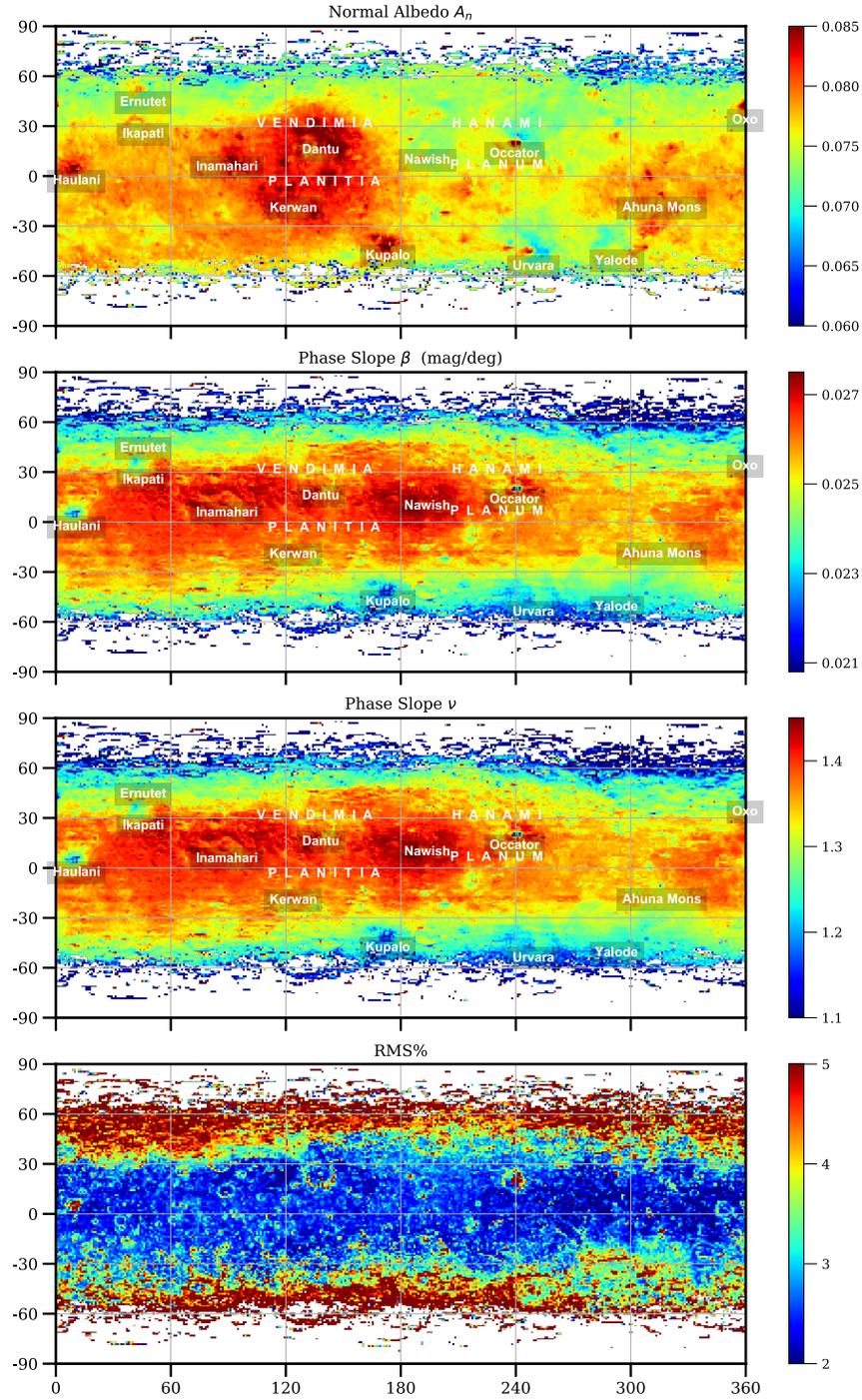

Figure 7. Maps of linear magnitude phase function model parameters with the Akimov disk-function in F2 (555 nm). The white areas at high latitudes are not mapped due to insufficient number of data points that satisfy our cutoff criteria. The map of phase slope $v$ and that of $\beta$ are identical except for a scaling factor. The normal albedo map and the $v$ map are displayed with the same scale bars as in Figure 10 of Schröder et al. (2017), and can be compared directly.

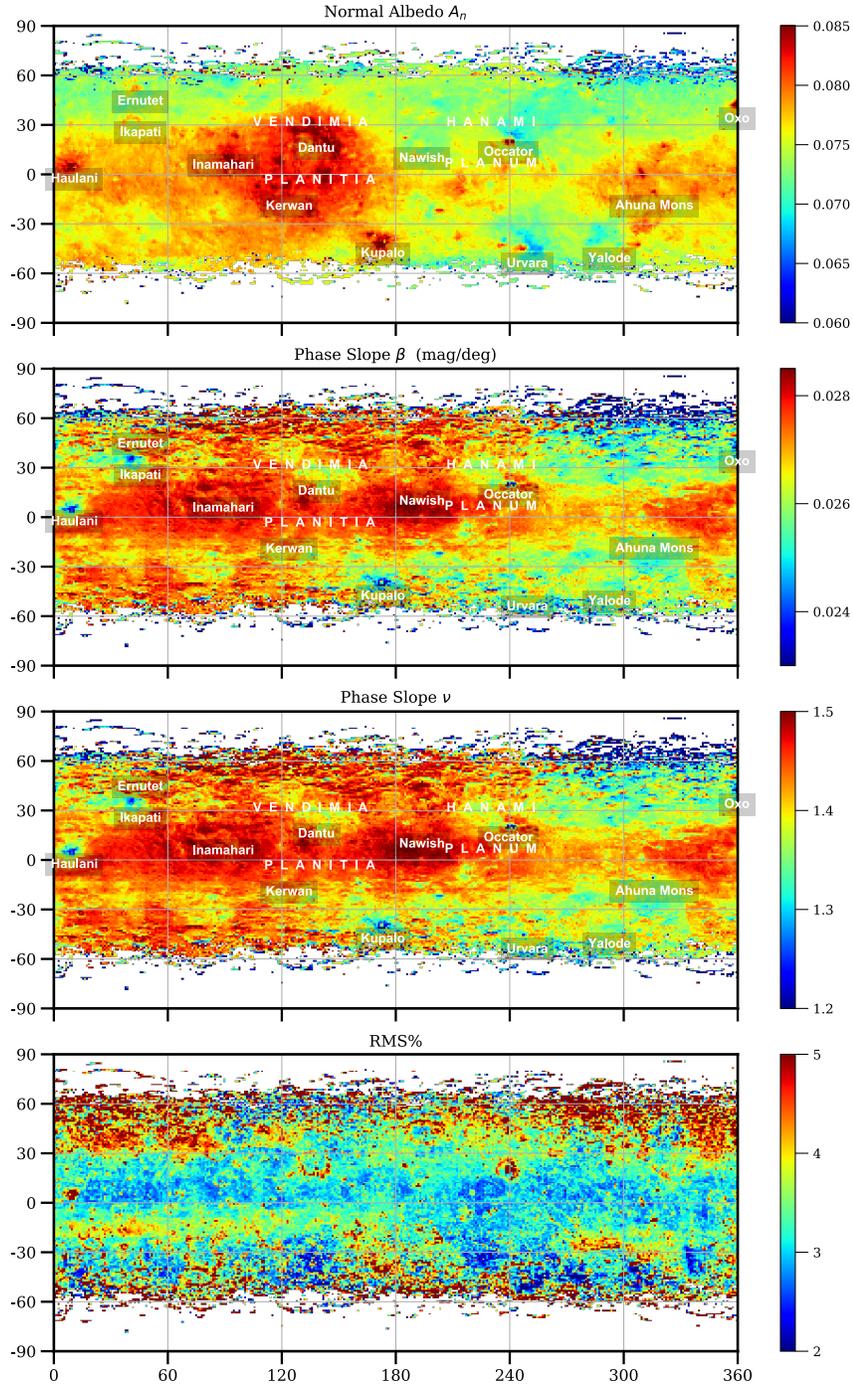

Figure 8. Same as in Figure 7 but derived with the LS disk-function model. The color scales in normal albedo map and the RMS map are the same as in Figure 7, but those of the phase slope maps are slightly different.

## 5.2. Hapke model mapping

As for the global photometric modeling, we set the opposition parameters with $B_0$=1.6 and $h$=0.06. The mapping results from the Hapke model with 1pHG are shown in Fig. 9. However, the Hapke model with 2pHG could not generate satisfactory maps: the maps of the SSA, *b* and *c* all contain many features that have obvious characteristics that are similar as in the map of maximum phase angle (Fig. 2), and therefore must be modeling artifacts. Because the modeling of 2pHG requires data at high phase angle to constrain both single-scattering phase function parameters, the lack of data at sufficiently high phase angle for the low latitude regions and the sharp boundaries between low and high latitude regions are the likely reasons that the 2pHG Hapke model did not work well for this mapping. We therefore did not include those maps in our discussion, except for the normal albedo maps.

Spatial variations are evident in all three free parameters, i.e., the SSA, the asymmetry factor, and the roughness. The SSA map shows overall similar characteristics as the normal albedo maps as derived from empirical models (Figs. 7 and 8), as well as the reflectance maps generate with traditional photometric correction approach (e.g., Fig. 7 in Schröder et al. 2017), suggesting that albedo variations dominate the reflectance variations on Ceres.

The asymmetry factor parameter $\xi$ shows a similar distribution as the phase slope maps derived from empirical models (Figs. 7 and 8). The strength of backscattering shows an overall anti-correlation with albedo for the low latitude region inside of ±30º latitude (Fig. 10), where relatively low albedo is associated with stronger backscattering and vice versa. This trend is similar to the general correlation between albedo and phase function in asteroids (Li et al., 2015), and is attributed to the fact that brighter, more transparent regolith grains tend to be more forward scattering (e.g., Souchon et al. 2011).

The roughness map also shows some degree of spatial distribution (Fig. 9). However, compared with the characteristic maps of photometric mapping data (Fig. 2), we immediately notice that it has some sawtooth pattern at about ±30º-45º latitude that is similar to the map of maximum phase angle distribution. Between these two latitudinal boundaries, the maximum phase angle is dominated by RC3 data; while outside these boundaries towards high latitude areas, the maximum phase angle is dominated by Survey data. Because the modeling of roughness is most sensitive to high phase angle data (Helfenstein 1988, see also Helfenstein et al. 1988), the existence of these features in the roughness map is certainly an artifact due to the sharp boundary in the maximum phase angle. In addition, the belt-like low roughness region centered at latitude +5º and extending east-west between longitude 20º and 100º (greenish in the map) is probably also a modeling artifact because it does not appear to be associated with any geological context. Other than those, there do not seem to be other identifiable artifacts in the map.

The roughness map does not show an overall correlation with albedo on the global scale (Fig. 10). However, on regional scales, there appear to be some correlations. The most prominent ones are the following. The relatively bright region along the northern side of Vendimia Planitia has relatively high albedo, weaker backscattering, and higher roughness. The Nawish crater region between the Vendimia Planitia and Hanami Planum has relatively low albedo, stronger backscattering, but also higher roughness than overall Ceres. On the other hand, the Hanami Planum, which has Occator crater located near just off the center, has relatively low albedo, moderate backscattering, but no obvious deviation in roughness from the surroundings. The range of roughness variations is about 5º. Although only slightly higher than the range of spectral variations of roughness (Fig. 3, Section 4.3), which we considered as modeling scatters, the spatial variations of roughness should be real as the patterns are clearly visible above the model scatter (background noise) in the map. The Hapke model SPPF and roughness mapping results suggest that the variations in phase slope over the surface of Ceres as revealed by empirical models (Figs. 7 and 8) are more likely dominated by SPPF than roughness, as previously reported by Schröder et al. (2017). Although the physical meaning or scale size of the Hapke roughness is not entirely understood, Hapke model mapping is still able to break the ambiguity between particle phase function and roughness and reveal the physical nature of these phase slope variations to some extent.

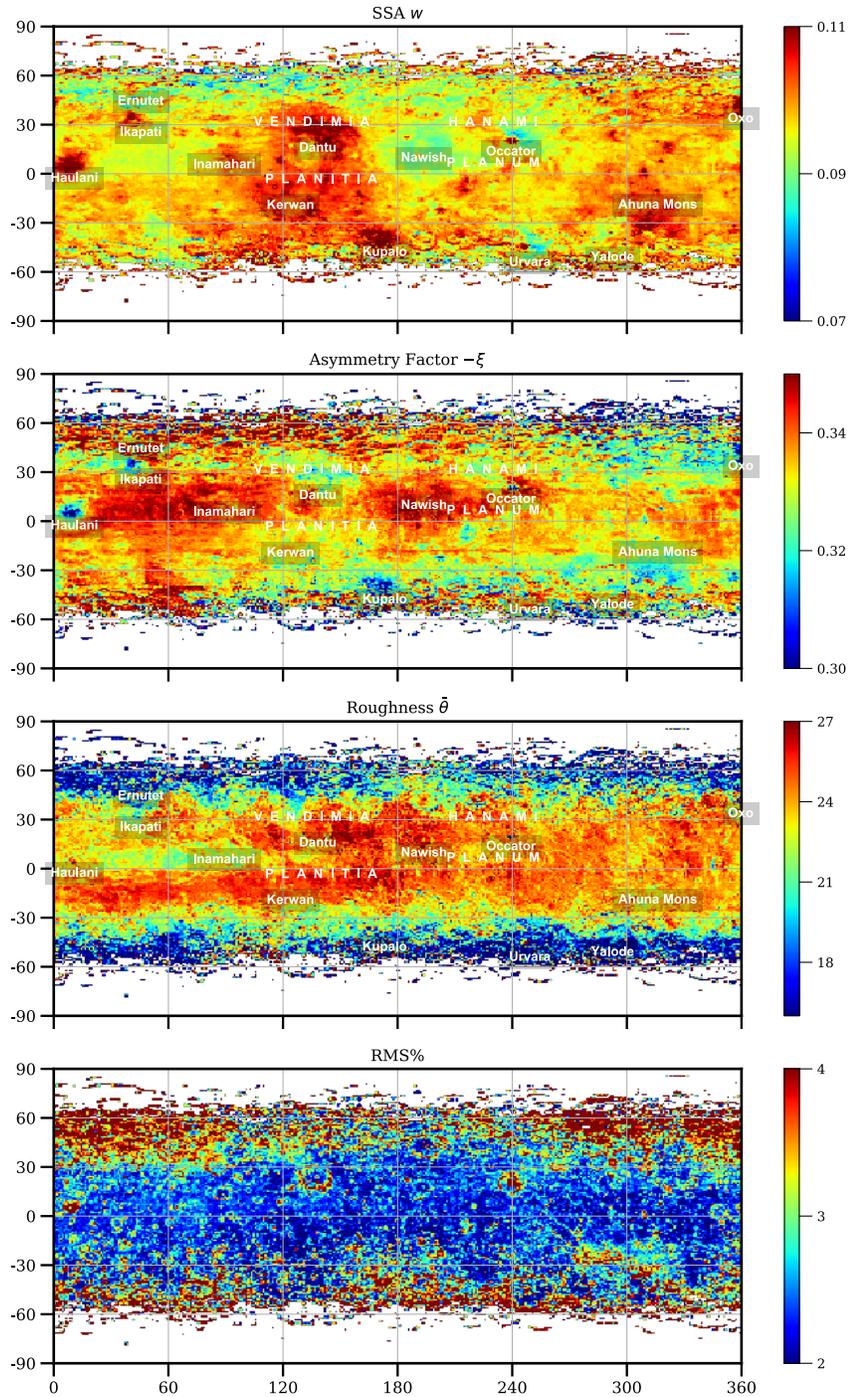

Figure 9. Maps of parameters and RMS of Ceres in F2 filter derived with the 5-parameter Hapke model. White areas are not mapped due to insufficient data points in the grid.

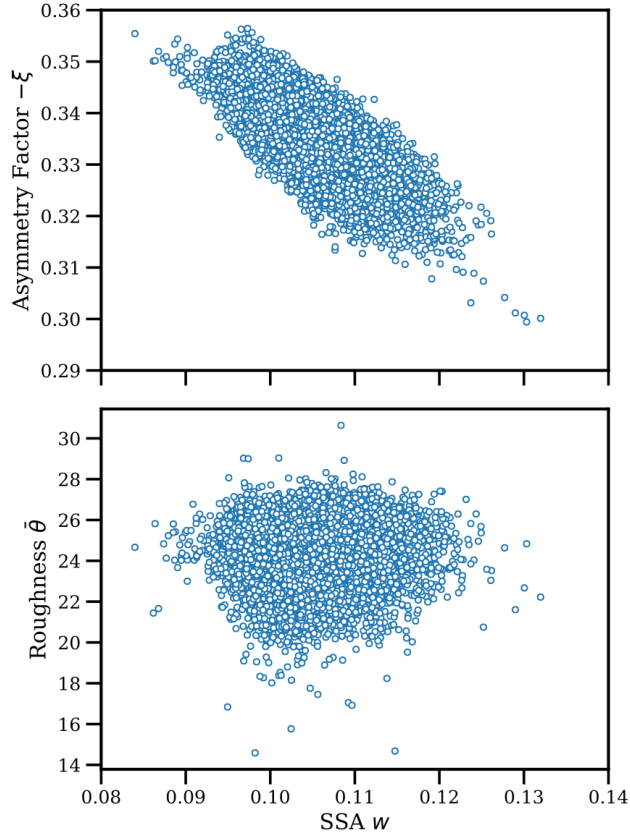

Figure 10. The upper panel shows the correlation between SSA and asymmetry factor (-ξ plotted) for the region between ±30° latitude in the F2 filter, where lower albedo corresponds to relatively stronger backscattering, and vice versa. The correlation coefficient is 0.72. The lower panel shows the overall lack of correlation between albedo and roughness parameter for the same area. We did not include high latitude regions outside of ±30° in this study because the photometric maps are not sufficiently reliable.

Compared with the global geologic map of Ceres (Williams et al., 2018a), the region where the highest roughness distributes appears to be associated with the ancient Vendemia Planitia basin underlying the young craters Dantu and Kerwan. Therefore, the high Hapke roughness in the Dantu crater region is associated with the fresh, possibly doubly excavated materials from relatively deep crust compared to other places on Ceres. Other young craters that are also associated with bright materials, such as Haulani and Occator etc., do not have this double-excavation setting and are not associated with high Hapke roughness. Furthermore, the Kerwan crater floor appears to be quite smooth in Survey and HAMO images (Williams et al., 2018b), but heavily cratered by small craters in LAMO (low-altitude mapping orbit) images with resolutions of about 35 m/pix. The high Hapke roughness could be associated with these small craters that are below the resolution of the data we used. In short, the areas on Ceres with high Hapke roughness, whatever its true physical interpretations are, could be related to Vendemia Planitia (Kerwan and Dantu) and their associated materials and geomorphology.

*5.3. Normal albedo*

The normal albedo maps derived from empirical models are shown in Figs. 7 and 8, and those derived from Hapke model with 1pHG and 2pHG are shown in Fig. 11. Despite the fact that the 2pHG Hapke model produced substantial artifacts in its individual parameter maps, the map of normal albedo is almost identical to that produced by the 1pHG Hapke model. This is because normal albedo is defined at 0º phase angle, it is minimally affected by the maximum phase angle of data used in modeling.

Comparisons among the normal albedo maps produced by all four models show an excellent agreement in the spatial distribution and the relative brightness scale almost everywhere down to the size of ~20 km, with only a slight difference in the north-west ejecta field of Occator crater as mentioned before. We consider these maps high fidelity. On the other hand, the absolute albedo scales of the maps produced by empirical models are lower than those of maps produced by Hapke models by about 24%. This is due to the fact that the empirical phase function that we adopted (Eq. 7) does not include the opposition effect, while the Hapke models do.

The histogram of the normal albedo map (after re-projected to sinusoidal projection) of Ceres shows a narrow, single-peak distribution (Fig. 12). The average normal albedo is 0.10 based on the normal albedo map, consistent with the geometric albedo of 0.096 from the global photometric modeling using the 1pHG Hapke model (Section 4). Note that the geometric albedo and average normal albedo of Ceres are expected to be close to one another, but not exactly the same for Ceres. The uncertainty for normal albedo estimate is similar to that of geometric albedo of about 0.006. The distribution of normal albedo is narrow, with a full-width-at-half-maximum of about 6% of the average, in excellent agreement with the previous observations from HST at about 30 km/pixel (Li et al., 2006). Generally, higher spatial resolution is able to bring up more extreme albedo features, if exist, to broaden the albedo distribution for planetary surfaces. Therefore, any features with extreme albedo on Ceres must be at scales smaller than a few km. The overall albedo distribution on Ceres is quite narrow, despite the existence of some small areas with extremely high albedo, such as Cerealia Facula (Li et al., 2016b, Schröder et al., 2017).

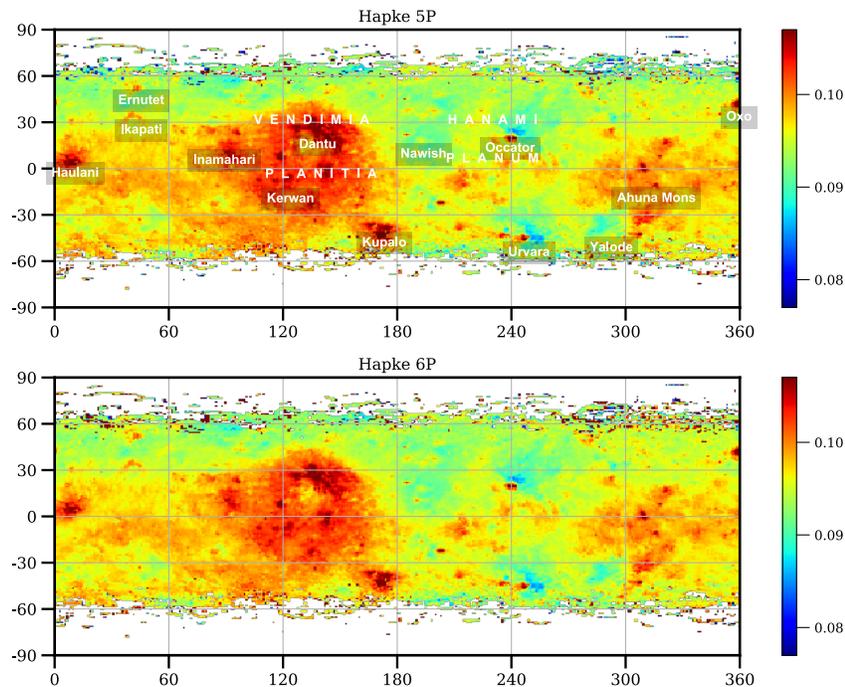

Figure 11. Normal albedo maps in F2 filter derived from the Hapke model with 1pHG (upper panel) and 2pHG (lower panel).

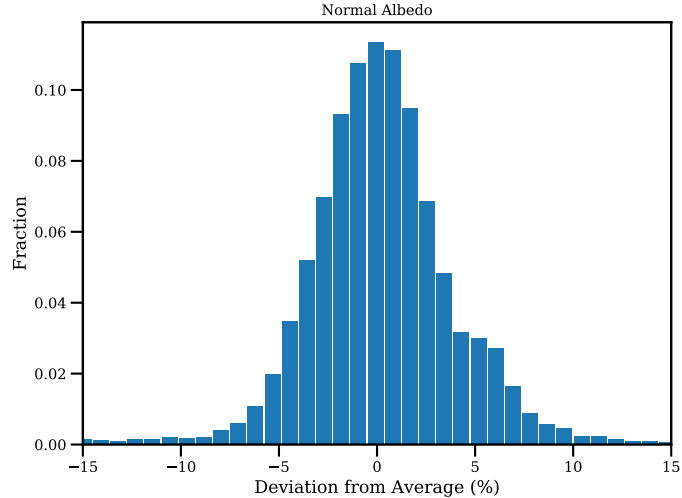

Figure 12. Ceres' normal albedo histogram in F2 filter.

The normal albedo, and by extrapolation the Bond albedo, of Ceres is rather uniform, and therefore the amount of absorbed solar energy therefore varies little over the globe. We zonally averaged the albedo map and repeated the depth-to-ice calculations described in Schorghofer (2016) and Prettyman et al. (2017). Changes in predicted depth-to-ice are less than 1%, and these albedo variations are too small to explain the hemispheric asymmetry observed in the hydrogen content (Prettyman et al., 2017).

*5.4. Wavelength dependence (color)*

In this section, we discuss the spatial variations of the wavelength dependence of the Hapke parameters on the surface of Ceres. Such variations manifest themselves as changes in the parameter maps from band to band. For this study, we generated various color composite maps by assigning the maps of the same parameter at various selected wavelengths, or the ratios of maps from different wavelengths, to red (R), green (G), and blue (B) channels. One color composite map we used assigns F5 (965 nm), F3 (749 nm), and F8 (438 nm) filters to RGB channels, respectively. This color composite is termed "enhanced color" in our work. The second color composite has the ratio of F5/F3, albedo in F3, and the ratio of F3/F8 in RGB, respectively, and is termed "ratio-albedo color", although it can be used for more parameters than just albedo. The third color composite uses the ratios of F5/F3, F2/F3, and F8/F3 for RGB, respectively, and we call it "ratio color". The enhanced color scheme is exactly what was adopted in the initial study of Ceres color properties by Nathues et al. (2016b), and similar to what was used by Schröder et al. (2017) where they replaced F3 with F2 (555 nm). The ratio color scheme is also the same as those used by Nathues et al. (2016b) and Schröder et al. (2017). We will use all three color-composite to study normal albedo maps, and the enhanced color only to study asymmetry factor and roughness maps. The meaning of these color composites will be discussed for each parameter.

The three-color composite maps of Ceres are shown in Fig. 13. The wavelength dependence of normal albedo is a spectrum in the usual sense. The enhanced color map corresponds to the color of the surface of Ceres in our common sense, but extends to UV (440 nm) and NIR (960 nm) with much exaggerated color stretch. Our enhanced color composite and ratio color composite appear to be similar to the previously reported maps by Nathues et al. (2016b) and Schröder et al. (2017), although with different stretches in color channels and different projections. We do not discuss them in detail here, and readers are referred to previous studies for the analysis and interpretations.

The enhanced color map of asymmetry factor is shown in Fig. 14. Overall the color variations in the map are bland, with only slight brightness patterns but not much color patterns. Some patterns, such as the sawtooth pattern at 120º to 300º longitude and -30º and 0º latitude, have similar distribution as the maximum phase angle map (Fig. 2) and must be artifacts. It is hard to say whether the slight magenta and greenish color contrast between west and east hemispheres is real or not, but given that its strength is similar to the sawtooth artifacts, they are likely artifacts. In addition, the horizontal line at about -20º latitude extending around the globe should also be an artifact due to its highly regular shape that does not appear to correlate with any geological features on Ceres. Compared to the asymmetry factor map in a single band (Fig. 9), the areas where backscattering is relatively enhanced in 20º to 120º longitude and 0º to +20º latitude, and in 160º to 230º longitude and 0º to 30º latitude disappears. The regions associated with some bright craters, such as Haulani and Kupalo where backscattering is relatively weak do not have much color variations.

From disk-integrated photometric modeling, we showed that the SPPF of Ceres has less backscattering towards longer wavelength (Fig. 3, Tables 1 and 2). This behavior is similar across the surface of Ceres, as suggested by the spectra of $\xi$ for a few areas that we checked (Fig. 14). To avoid possible artifacts in latitudinal direction because of the different ranges of scattering geometry (especially the maximum phase angle, Fig. 2), the features we checked are between 0º and +30º latitude. They all have similar overall slope across the visible wavelengths of the FC filters, despite the scatters at some wavelengths, although the absolute values are different, with bright craters such as Haulani relatively less backscattering than dark areas such as the dark ejecta of Occator crater. In summary, the color map of asymmetry factor suggests that its wavelength dependence does not vary much over the surface.

Similar to the asymmetry factor, the roughness parameter does not show much wavelength dependence over the whole surface of Ceres either (Fig. 15). The band with light magenta color at 0º to 15º latitude over the full longitude, as well as the sawtooth shaped patterns, are all artifacts, again due to the distribution of maximum phase angle (Fig. 2). As we discussed before, roughness should not depend on wavelength. The roughness spectra of five locations on Ceres all show similar shapes as the global average roughness parameter as shown in Fig. 3.

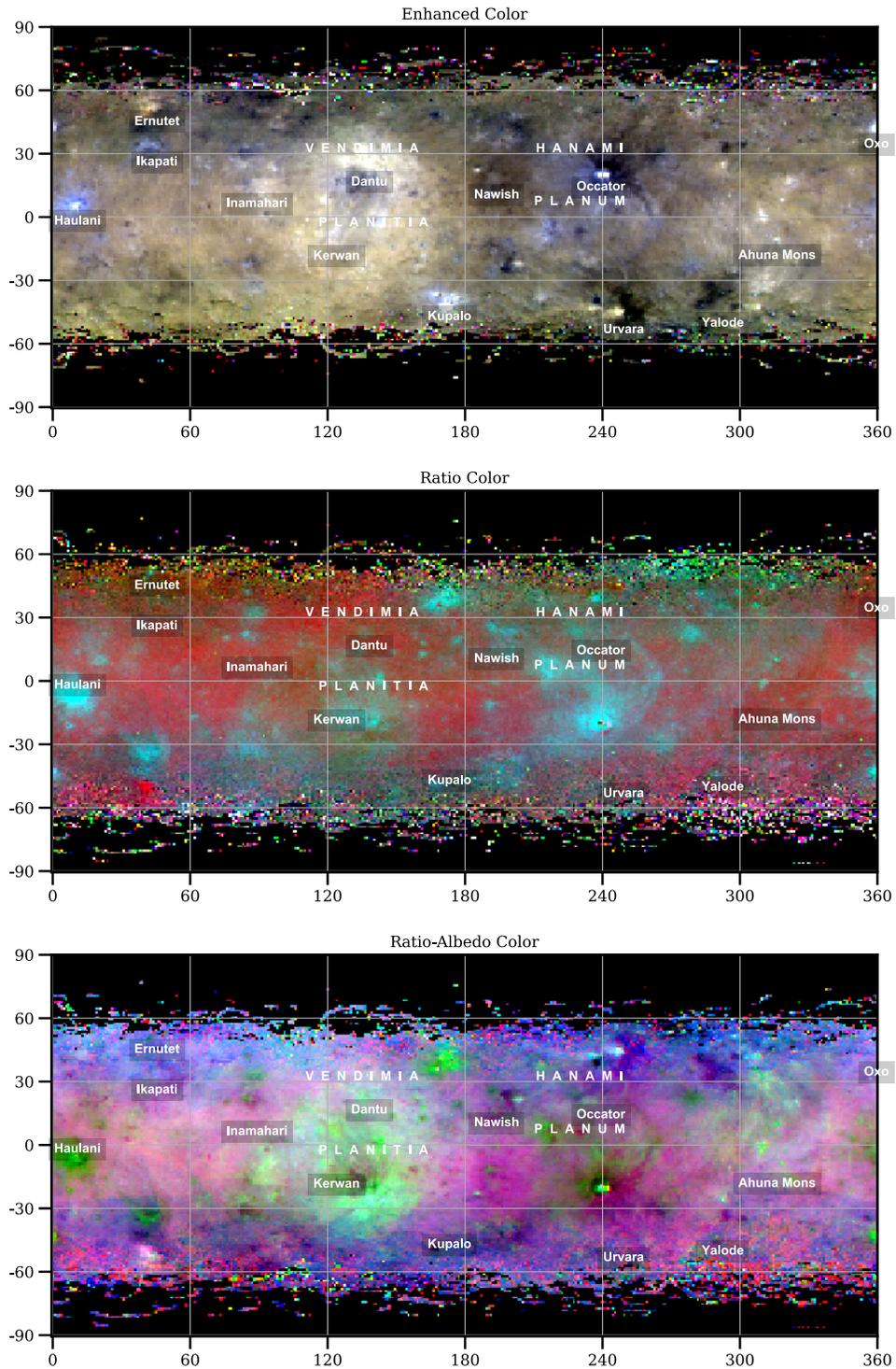

Figure 13. Color composite maps of Ceres: enhanced color map (upper panel), ratio-albedo color map (middle panel), and ratio color map (lower panel). See text for the color assignment scheme and description of these color maps. Some major geological features are marked in the maps right above the corresponding labels.

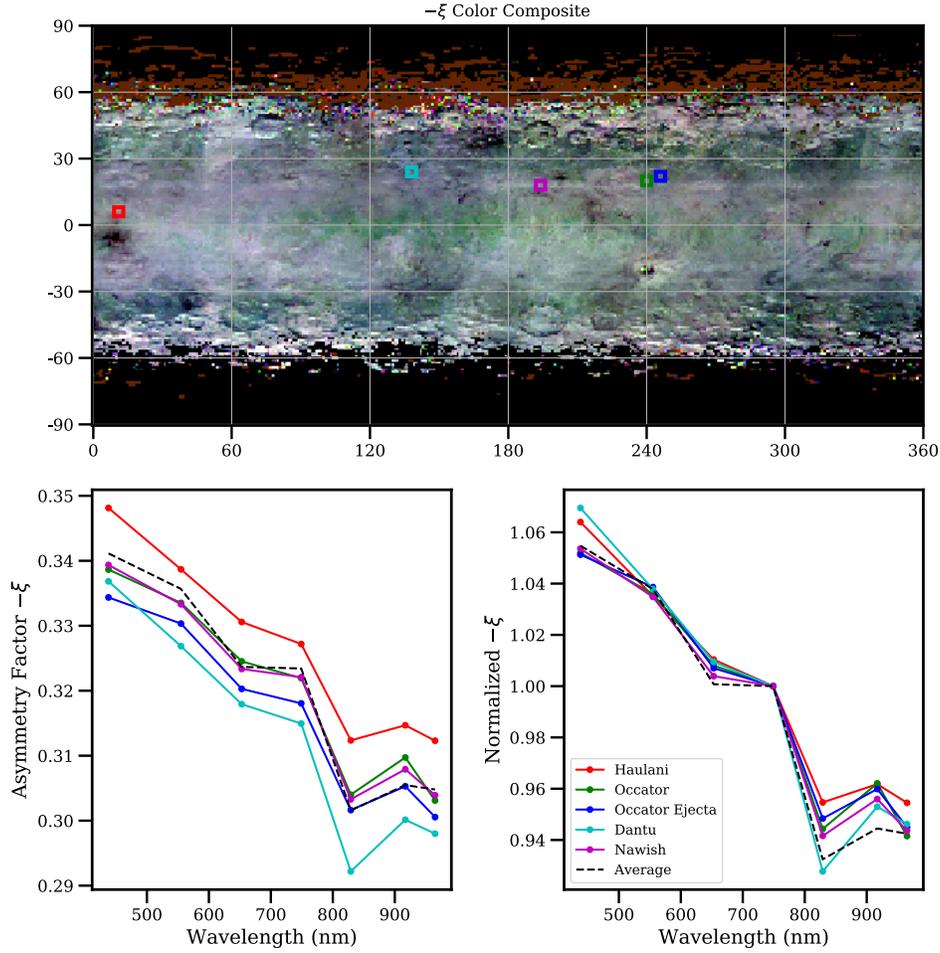

Figure 14. Enhanced color map of the asymmetry factor $\xi$ (upper panel), and the spectral plot of $\xi$ for selected regions (lower panels). The bottom left panel plots the spectra directly and the bottom right panel plots the same spectra normalized to the values at 750 nm. Note that we used -$\xi$ in the map and plots. The plot uses average values inside 4º×4º boxes centered at the features as marked in the map. The color variations in this color map are mild.

## 6. Discussion

### 6.1. Forward scattering

As discussed in Section 4.2, the phase function of Ceres is better described with a 2pHG, and 1pHG results in a systematic bias in the model. The comparisons with other asteroids previously analyzed with spacecraft data suggested that the SPPF of Ceres is so far unique, except for perhaps (21) Lutetia (Table 3, Fig. 16). Asteroids (2867) Šteins was studied with 1pHG and 2pHG, as well as 3-parameter HG function (3pHG) where there are two separate parameters for backward and forward scattering terms in Eq. 3, and 1pHG was able to fit the phase function well (Spjuth et al., 2012). For Lutetia, the disk-resolved data at phase angles 0º-95º could be well modeled with a 1pHG, although its disk-integrated phase function at phase angles 0º-160º needed a 2pHG to model (Masoumzadeh et al., 2015). As shown in Fig. 16, the two models for Lutetia results in the same

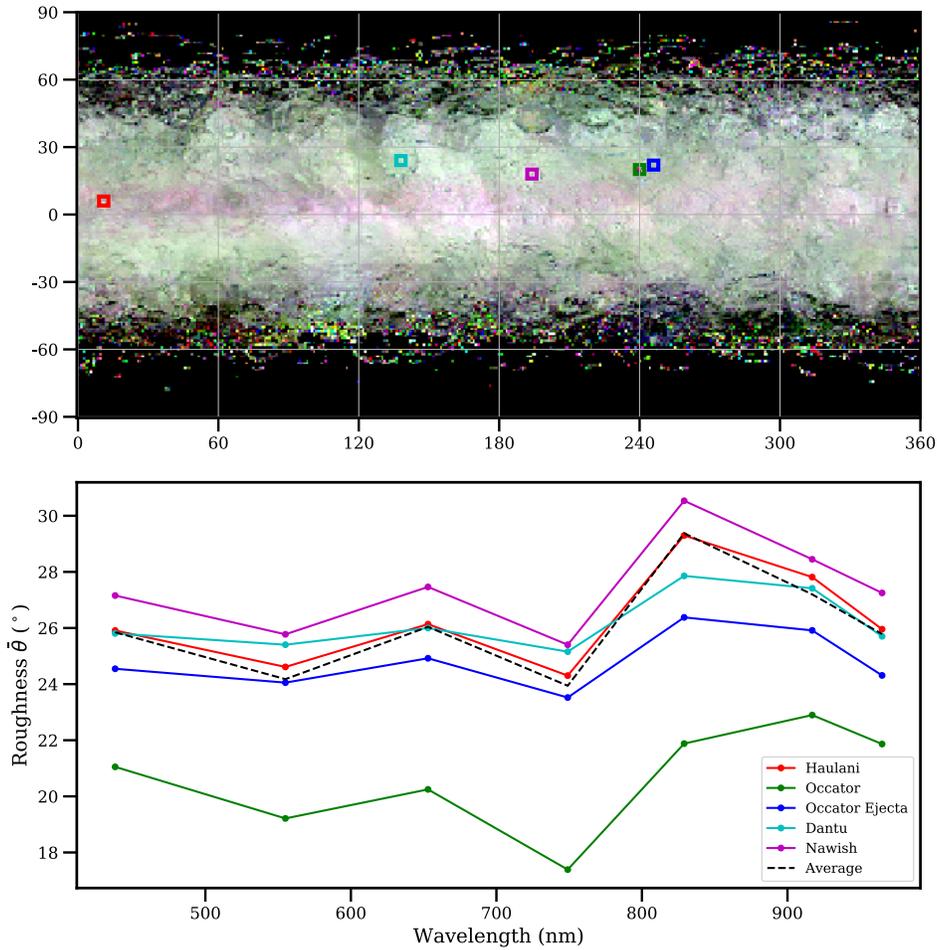

Figure 15. Enhanced color map of roughness (upper panel) and roughness "spectra" of selected regions on Ceres (lower panel). The horizontal band in light magenta color along the equator, as well as the sawtooth patterns are all likely artifacts due to the change in maximum phase angles for the data used in the modeling. The plot uses average values inside 4°×4° boxes centered at the features as marked in the map. No wavelength dependence of roughness is evident across Ceres' surface.

SPPF at phase angles less than about 60º, then starts to diverge towards higher phase angles. Hasselmann et al. (2016) analyzed the Baetica region on Lutetia and found an overall consistent but slightly more backscattering results in its photometric properties than the global average. (4) Vesta (Li et al. 2013) could be modeled with a 1pHG without systematic bias. Domingue et al. (2002) used 2pHG to model (433) Eros with the NEAR/MSI data, but found that the forward scattering term is not needed, suggesting that the phase function can be well fitted by 1pHG. Li et al. (2004) were able to fit the phase functions of Eros in the visible wavelengths with 1pHG. (253) Mathilde was modeled with both 1pHG and 3pHG, and the 1pHG fitted data well (Clark et al., 1999). (243) Ida (Helfenstein et al., 1996), (951) Gaspra (Helfenstein et al., 1994), and (25143) Itokawa (Li et al., 2018) were all fitted well with 1pHG, although these data were either much poorer in quality than those from later missions or have relatively narrower coverages in phase angle. We note that the disk-resolved data at phase angles >100º are available only for Steins (up to 130º), and the disk-integrated data beyond 100º are available only for Lutetia (up to 160º),

Mathilde (up to 130º), and Ida (up to 110º). Even though we restrict the comparisons of SPPFs within the phase angles where data are available, Ceres still displays a distinctly different scattering behavior at phase angles around 90º (Fig. 16 insert). The normalized SPPF of Ceres flattens out and turns up with respect to phase angle at around 90º, while the SPPFs of other asteroids have much steeper decreases with respect to phase angle. Such difference could be an indication that the forward scattering of Ceres starts at relatively lower phase angles.

Are the differences between the SPPFs of those asteroids statistically meaningful? For most spacecraft targets (perhaps except for the earliest flyby targets Gaspra and Ida), it is reasonable to consider that the uncertainties in the shapes of their SPPFs are comparable to that of Ceres as we determined in our work. As shown in Fig. 4, the difference between the 1pHG and 2pHG best-fit models are statistically significant and reliable for Ceres. Furthermore, the difference between the 1pHG of other objects and the 2pHG of Ceres and Lutetia near 90º phase angle (Fig. 16) is comparable to the difference between the 1pHG and 2pHG of Ceres (Fig. 6). Therefore, we consider that the differences between Ceres and other objects (except for Lutetia 2pHG) as shown in Fig. 16 are statistically meaningful.

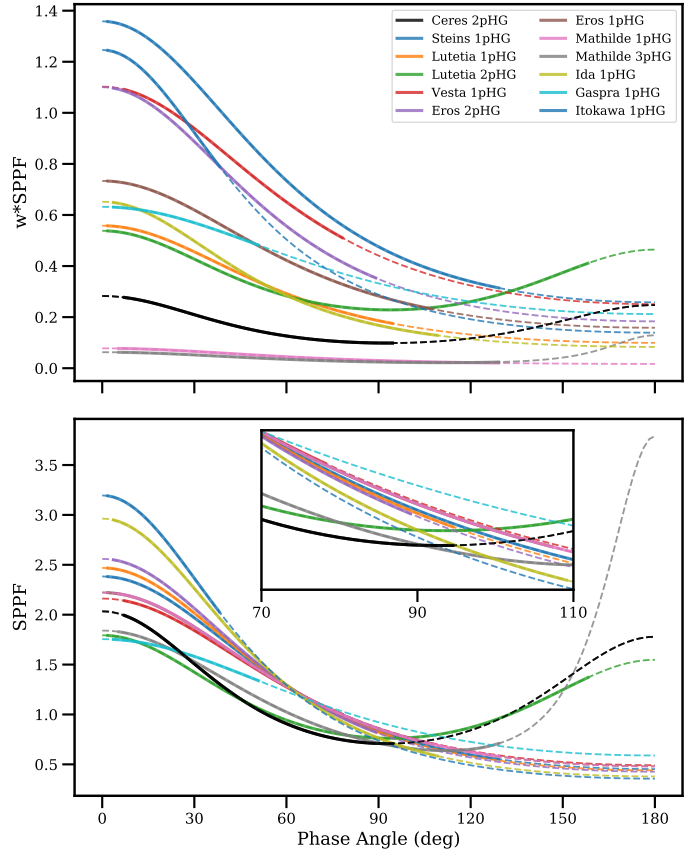

Figure 16. The comparison between the best-fit SPPF for Ceres and those of other asteroids listed in Table 3. Upper panel plots the product of SSA and SPPF with respect to phase angle; lower panel is the normalized SPPF with its integral over all $4\pi$ solid angle to be unity; and the insert in the lower panel shows the details at moderate phase angles between 70º and 110º. The solid part of each curve is where data are available to constrain the SPPF models for the corresponding objects, and the dashed part is extrapolated based on the SPPF model.

The fact that Ceres' regolith starts to scatter more light from relatively lower phase angle than that of other asteroids is intriguing. We can gain some insights about the physical characteristics of Ceres regolith from its phase function based on relevant laboratory work of planetary surface simulants (McGuire and Hapke, 1995; Souchon et al., 2011). In the plot of *b* vs. *c* as measured from the laboratory, Ceres is in a location between the grains with medium and low densities of internal scatterers (Fig. 17). For other asteroids with their SPPFs plotted in Fig. 16, the fact that a 1pHG is able to describe their SPPF means that backscattering dominates, and their *c*-parameters are likely distributed in the upper region in the "Hockey-Stick" plot (Fig. 17). Therefore, the regolith grains of Ceres are expected to have rough surfaces and contain relatively fewer internal scatterers compared to those on other asteroids.

What might cause such differences in the physical properties of regolith grains on Ceres compared to other asteroids? The primary difference between Ceres and other asteroids on the global scale is probably the ubiquitous phyllosilicates distribution and the relatively high abundance of carbonates (De Sanctis et al., 2015; Ammannito et al., 2016). For those asteroids listed in Table 3, the only other one that could have a similar composition as Ceres is Mathilde. However, neither the 0.7 μm nor the 2.8 μm feature that are commonly associated with hydration in phyllosilicates is evident in the spectrum of Mathilde, whose near-IR spectrum appears to be consistent with a sample of Murchison heated to 900º C (Binzel et al., 1996; Rivkin et al., 1997). Modeling suggested that the average temperatures at and near the surfaces of Ceres are never expected to exceed 300 K (e.g., McCord and Sotin, 2005; Castillo-Rogez and McCord, 2010; Neveu et al., 2015; Formisano et al., 2016a; b). In addition, ample evidence suggests that water ice, water of hydration, or even liquid water is present on or close to the surface of Ceres (e.g., Combe et al., 2016; Ruesch et al., 2016; Sizemore et al., 2017; Prettyman et al., 2017; Schmidt et al., 2017; Nathues et al., 2017, etc.). Therefore, the regolith of Ceres is aqueously altered, never heated, and rich in water ice and/or hydration. Interestingly, laboratory experiments showed that Mars soils analogs become more forward scattering after wetting by a few percent of water or water ice, and even after completed drying up (Pommerol et al., 2013). The SPPF of Ceres is also compatible with that of the phyllosilicate sample nontronite in the visible as measured in the laboratory (Pilorget et al., 2016). Therefore, the water-rich and aqueously altered composition of Ceres might be associated with its relatively strong forward scattering compared to other asteroids imaged by spacecraft so far. We should probably expect similar behaviors for other asteroids of similar compositions.

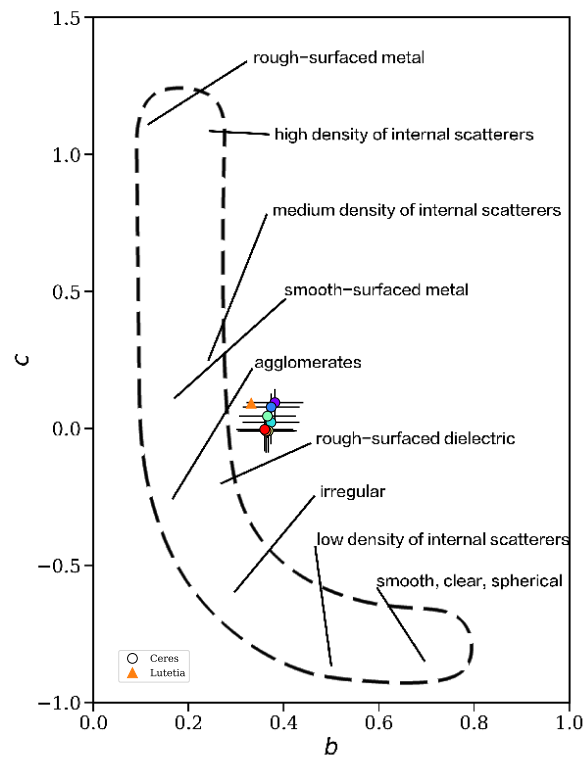

Figure 17. The "Hockey-Stick" plot of the 2pHG. The color circle symbols with error bars are the *b* and *c* parameters of Ceres from 438 nm filter (purple symbol) to 965 nm (red symbol), as listed in Table 1. The triangle symbol is for the 2pHG parameters of Lutetia (Masoumzadeh et al., 2015). The physical interpretations of the regions in this *b-c* plot follows McGuire and Hapke (1995).

*6.2. Spatial variations in phase function*

Empirical modeling shows that the slope of the surface phase function varies across the surface of Ceres (Section 5.1, Figs. 7 and 8, and Schröder et al., 2017). The phase function combines the effects of opposition effect, SPPF, and roughness. While it is relatively certain that the variations in Vesta's surface phase function are likely caused by roughness associated with various geological settings (Schröder et al., 2013a), it is not clear that geological settings are the predominant causes for such variations in the case of Ceres. Our photometric mapping with the Hapke model suggests that it is likely the SPPF, rather than the roughness parameter, that dominates such variations.

The spatial variations of $\xi$ across Ceres surface appear to be correlated with albedo (Figs. 9 & 10). For the range of SSA of 0.09 – 0.12, the corresponding variations in $\xi$ is about -0.35 to -0.31 (Fig. 10). The SPPF is generally determined by the physical characteristics of regolith grains (McGuire and Hapke, 1995; Souchon et al., 2011). We consider that the most likely cause for these variations should be the transparency of regolith grains, where grains with relatively higher transparency increases the albedo, and make the scattering function relatively more isotropic (less backscattering). Because the correlation between albedo and phase slope is commonly found for asteroids (Li et al., 2015), it seems prudent that, for the interpretations of any phase slope variations, we should first check whether there is any correlation with albedo. If such correlation exists, one must first estimate how much variation in phase slope might be caused by the variations in the SPPF, before attributing phase slope variations to roughness variations.

Based on these principles, we went back and checked our interpretations for the photometric variations of Ceres as presented here, as well as those for Vesta as presented by Schröder et al. (2013a). For Ceres, the variations in phase slope are in general correlated with albedo (Figs. 7 and 8), and we show that most of these variations are caused by variations in SPPF (Figs. 9 and 10). The variations in roughness are concentrated in local areas, but generally minimal on a global scale. For Vesta, on the other hand, the areas where there are prominent variations in phase slope generally do not show prominent variations in normal albedo, or show a correlation with normal albedo that are opposite to the general albedo-phase slope correlation aforementioned. Those areas include the ejecta field of Cornelia crater and the southern floor of Numisia crater (Fig. 13 in Schröder et al., 2013a), the ejecta field of Tuccia crater, the debris field in the southern part of Antonia crater floor, and the wall of Mariamne crater (Fig. 14 in Schröder et al., 2013a). Therefore, the interpretation that the phase slope variations for those areas are due primarily to roughness but not SPPF is justified.

*6.3. Phase reddening (wavelength dependence of light scattering)*

At a first glance, the phase reddening behavior of Ceres does not seem to be special when compared to other objects (Section 4.2). However, a detailed analysis offers us some insights into the phase reddening as well as the physical properties of Ceres regolith grains.

Phase reddening is equivalent to wavelength dependence of surface phase function, or specifically, shallower phase slope (less backscattering) towards longer wavelengths. For asteroids with a silicate composition, such as Vesta, Eros, and Itokawa, it has long been noticed that their asymmetry factors, $\xi$, only show a weak dependence on wavelength (Li et al., 2013; Clark et al., 2002; Li et al., 2004; Kitazato et al., 2008; Li et al., 2018), whereas their spectra show a general red slope outside the 1-µm and 2-µm mafic bands (e.g., Reddy et al., 2011; Murchie and Pieters, 1996; Abe et al., 2006). In the Hapke model framework (Eq. 1), increased albedo at longer wavelengths increases the multiple scattering term, $H(\mu_0, w)H(\mu, w) - 1$, relative to the single scattering term. Therefore, it is generally considered that the increase of multiple scattering towards longer wavelengths causes phase reddening, while the SPPF should not have much effect (Muinonen et al., 2002). The deepening of the 1-µm and 2-µm bands with increasing phase angle for Vesta (Reddy et al., 2011; Longobardo et al., 2014) is also consistent with this hypothesis. In addition, recent laboratory studies suggested that small-scale surface roughness could also play a role in determining the characteristics of phase reddening (Beck et al., 2012; Schröder et al., 2014b).

Compared to silicate composition asteroids, Ceres has a much lower albedo, and displays a flat spectrum in the visible and near-IR spectral range (cf. Rivkin et al., 2011; Nathues et al., 2015b). Multiple scattering is thus expected to be much lower than for those asteroids and should not change much with wavelength. On the other hand, the SPPF of Ceres clearly shows a trend of weaker backscattering towards longer wavelengths (Fig. 3). Therefore, phase reddening of Ceres is not likely controlled by multiple scattering, but more likely by single scattering and/or small-scale roughness.

If single scattering is the cause of phase reddening for Ceres, what could cause the wavelength dependence of SPPF for Ceres? Laboratory studies suggested that SPPFs are affected by, among other factors, the characteristics of surface and/or internal scatterers of grains (Souchon et al., 2011). Pilorget et al. (2016) analyzed the wavelength dependence of the SPPFs of the laboratory samples of basalt, olivine, phyllosilicate, and carbonate, and showed similar behavior in their carbonate sample (magnesite) in the visible, where more forward scattering (decreasing *c*) and less prominent anisotropic lobe (decreasing *b*) appear with increasing wavelength. The SPPFs of all other samples have different types of wavelength dependence. Based on the SEM imaging and the absorptivity analysis of their samples, Pilorget et al. (2016) suggested that the interaction of light with the surface structure of scattering grains, such as the roughness and the µm scale particles covering the surface, causes the wavelength dependence of their scattering behaviors. Therefore, we hypothesize that the regolith grains on the surface of Ceres either contain a considerable fraction of µm-sized or smaller grains, as suggested by Vernazza et al. (2017), or are strongly affected by those small-scale surface or internal scatterers, such as defects, impurities, or voids. The scattering efficiency of these small scatterers in the visible decreases with wavelength, and so the grains tend to be more transparent and less backscattering at longer wavelengths where the internal scatterers become less significant. Based on this hypothesis, the similar wavelength dependence of the asymmetry factor across the whole surface of Ceres (Section 5.4, Fig. 14) indicates that the properties of internal scatterers in Ceres regolith grains do not vary spatially. On the other hand, those other asteroids whose SPPFs do not depend on wavelengths may have regolith grains that are larger in size, or contain internal scatterers a few µm or larger.

The small grain size in Ceres regolith is consistent with the measured thermal inertia of the surface as well as with vapor diffusivity requirements inferred from nuclear spectroscopy. Earth-based observations indicate the thermal inertia of Ceres is about 15 [J m$^{-2}$ K$^{-1}$ s$^{-0.5}$] (Rivkin et al., 2011). Recent laboratory measurements by Sakatani et al. (2018) confirm extremely low thermal conductivity values for small grain size and high porosity. For the thermal environment on Ceres specifically, the thermal inertia value is consistent with particle sizes well below 100 µm (Schorghofer, 2016). The existence of near surface water ice at mid- and high-latitudes (Prettyman et al., 2017) also requires small grain size because this ice is lost to space by diffusion through the porous surface, with smaller pore sizes leading to slower diffusion. Models of ice loss suggest that the shallow depths to ice are best matched if the grain size (which affects pore size) is assumed to be around 1 µm (Prettyman et al., 2017).

## 7. Conclusions

In this work, we performed a detailed modeling of the global average photometric properties of Ceres, as well as a mapping of the photometric model parameters at seven visible wavelengths from 438 nm to 965 nm using the color images collected by Dawn Framing Camera during the Rotational Characterization 3 and Survey mission phases. The data have a pixel scale of 1.3 km/pix and 0.45 km/pix, respectively over the two phases, and cover almost the entire surface of Ceres at the full range of incidence and emission angles, and from phase angles 7º to 95º. We used the empirical models with the Lommel-Seeliger and the Akimov disk-function models coupled with an exponential phase function model, as well as the Hapke models with both a 1-parameter and a 2-parameter Henyey-Greenstein function in our modeling. We summarize the main findings as follows.

With no data at phase angles less than 7º, we had to fix the opposition parameters at $B_0$=1.6 and $h$=0.06 during our Hapke modeling process. The best-fit Hapke model parameters at seven wavelengths together with their formal statistical uncertainties are listed in Tables 1 and 2. The Hapke model fitting to the reflectance factor data suggested that the scattering properties of Ceres at phase angles from 60º to 95º are better modeled by a 2-parameter Henyey-Greenstein single-particle phase function, while a 1-parameter Henyey-Greenstein function results in a statistically significant, systematic trend with respect to phase angle. The phase function of Ceres shows a decreasing slope with increasing wavelength, suggesting phase reddening in visible wavelengths that is consistent with previous ground-based observations (Tedesco et al., 1983; Li et al., 2016b), as well as the observations by the Dawn VIR instrument (Ciarniello et al., 2017; Longobardo et al., 2018). The roughness parameters of Ceres have no dependence on wavelength, as expected for a geometric parameter.

In order to study the spatial variations in the photometric characteristics of Ceres regolith across the globe, we divided the surface into a grid with 1º in both latitude and longitude and performed photometric modeling to each grid independently in order to map out the best-fit photometric parameters for both the empirical models and the Hapke models. Spatial variations are evident in all the photometric parameters that we fitted, including normal albedo, phase slope, single-scattering albedo, single-particle phase function, and the Hapke roughness parameter. The albedo maps and various color composite maps produced this way are consistent with previous work following the traditional approach of photometric correction and mosaicking (Nathues et al., 2016b), or from a similar approach as in this work with empirical models only (Schröder et al., 2017). As also suggested by those previous work, the albedo and color show a clear correlation with geological features, where bright and blue regions are generally associated with geologically young craters. On the other hand, unlike Vesta (Schröder et al., 2013a), the spatial variations in the phase slope across the surface of Ceres seem to be dominated by the variations in single particle phase function rather than in roughness. The correlation between geomorphological features and roughness as observed on Vesta (Schröder et al., 2013a) does not exist on Ceres, although some correlation on large scale geomorphological features of hundreds of kilometers in size appears to be evident and might be attributed to particular geological processes. An example is the slightly higher roughness in the Vendimia Planitia region, where the surface could have been impacted and excavated multiple times, and the terrain appears to be heavily crated by small craters of less than 100 m in size. Mapping of photometric parameters in all color bands suggests weak dependence of their spatial variations on wavelengths.

The relatively strong light scattering from Ceres at moderate phase angles around 90º in the visible wavelengths compared to other asteroids previously studied using spatially resolved spacecraft data could be an indication of distinct scattering behavior of Ceres regolith grains. Based on previous laboratory measurements (McGuire and Hapke, 1995; Souchon et al., 2011), the modeled 2-parameter Henyey-Greenstein single particle phase function of Ceres could suggest regolith grains with medium to low densities of internal scatterers. The wavelength dependence of the single particle phase function might be an indication of sub-micron grains or internal scatterers for Ceres regolith, consistent with thermal modeling results (Schorghofer, 2016). Such distinct characteristics of light scattering might be associated with the unique surface composition of Ceres that is dominated by ubiquitous phyllosilicates from aqueous alteration, relatively high abundance of carbonates, and the mild heating history on or near the surface.

## Acknowledgements

We thank the Dawn Operations Team for the development, cruise, orbital insertion, and operations of the Dawn spacecraft at Ceres. The Framing Camera project is financially supported by the Max Planck Society and the German Space Agency, DLR. J.-Y. Li is supported by a subcontract from University of California at Los Angeles under the NASA Contract #NNM05AA86 Dawn Discovery Mission, as well as the Solar System Exploration Research Virtual Institute 2016 (SSERVI16) Cooperative Agreement (#NNH16ZDA001N), SSERVI-TREX. Part of this work was carried out at the Jet Propulsion Laboratory, California Institute of Technology, under contract to NASA. J.-Y. Li is grateful to Bethany L. Ehlmann for her insightful suggestions in the interpretation of the results. We thank Paul Helfenstein and an anonymous reviewer for their constructive suggestions that helped improve this manuscript. This research made use of a number of open source Python packages (in arbitrary order): Astropy (Astropy Collaboration et al. 2013), Matplotlib (Hunter 2007), SciPy (Jones et al., 2001), IPython (Pérez and Granger, 2007), Jupyter notebooks (Kluyver et al., 2016).

Table 3. List of the SPPF model parameters for asteroids imaged by spacecraft.

| Object | Type | Range of Phase Angle (º) and instrument | HG functional form for SPPF | SSA | $g$ (1pHG) or $b$ (2pHG) or $g1$ (3pHG) | $c$ (2pHG) or $f$ (3pHG) | $g2$ (3pHG) | Wavelength (nm) | Reference |
|---|---|---|---|---|---|---|---|---|---|
| (1) Ceres | C | 7 – 95 (Dawn/FC) | 1p, 2p (better) | 0.14±0.04 | 0.37±0.06 | 0.081±0.06 | | 555 | This work |
| (2867) Šteins | E | 0 – 130 (Rosetta/OSIRIS) | 1p (best), 2p, 3p | 0.57±0.05 | -0.28±0.02 | | | 630 | Spjuth et al. (2012) |
| (21) Lutetia | X | 0.5 – 95 (resolved) (Rosetta/OSIRIS) | 1p | 0.226±0.002 | -0.28±0.01 | | | 632 | Masoumzadeh et al. (2015) |
| | | 0.5 – 156 (integrated) (Rosetta/OSIRIS) | 2p | 0.30 | 0.33 | 0.095 | | | |
| (4) Vesta | V | 8 – 81 (Dawn/FC) | 1p | 0.51±0.08 | 0.51±0.08 | | | 555 | Li et al. (2013) |
| (433) Eros | S | 54 – 89 (NEAR/MSI) 4 – 58 (Ground), integrated | 2p | 0.43±0.02 | 0.29±0.02 | 1.0±0.11 | | 550 | Domingue et al. (2002) |
| | | 54 – 108 (NEAR/MSI) 1 – 57 (Ground), resolved | 1p | 0.33±0.03 | -0.25±0.02 | | | 550 | Li et al. (2004) |
| (253) Mathilde | C | 42 – 136 (NEAR/MSI) 1 – 17 (Ground) | 1p | 0.035±0.006 | -0.25±0.04 | | | 700 | Clark et al. (1999) |
| | | | 3p | 0.034±0.006 | -0.27±0.04 | 0.24±0.09 | 0.66±0.01 | | |
| (243) Ida | S | 20 – 60, 110 (Galileo/SSI) 0.6 – 21 (Ground) | 1p | 0.22±0.02 | -0.33±0.01 | | | 550 | Helfenstein et al. (1996) |
| (951) Gaspra | S | 33 – 51 (Galileo/SSI) 2 – 25 (Ground) | 1p | 0.36±0.07 | -0.18±0.04 | | | 550 | Helfenstein et al. (1994) |
| (25143) Itokawa | S | 0 – 39 (Hayabusa/AMICA) | 1p | 0.39 | -0.35 | | | 550 | Li et al. (2018) |

Notes:
The SPPF parameter values for 1pHG and 2pHG have been adopted to be consistent with Eqs. 2 and 3, respectively, in this work. The parameters for 3pHG (for asteroid 253 Mathilde) are as in Clark et al. (1999) following their adopted model form.